\documentclass[twocolumn]{aastex63}
\usepackage{amsmath}
\usepackage{xcolor}

\usepackage{ae,aecompl}
\usepackage{graphicx}	
\usepackage{amsmath}	
\usepackage{graphicx}
\usepackage{bm}
\usepackage{graphics}
\usepackage{afterpage}
\usepackage{float}
\usepackage{subfigure}
\usepackage{rotating}
\usepackage{multirow}
\usepackage{tabularx}
\usepackage{booktabs}
\usepackage{multirow}
\usepackage{fancyhdr}

\usepackage{theorem}
\usepackage{moreverb}
\usepackage{euscript}
\usepackage{psfrag}
\usepackage{slashed}
\usepackage{mathtools}
\usepackage{makecell}
\usepackage[flushleft]{threeparttable}
\usepackage{lineno}

\newcommand{\update}{\textcolor{black}}

\accepted{March 21, 2022}

\shorttitle{H$I$ absorption and the Galactic Center Excess}
\shortauthors{Pohl, Macias, Coleman, and Gordon}

\begin{document}

\title{Assessing the Impact of Hydrogen Absorption on the Characteristics of the Galactic Center Excess}

\author[0000-0001-7861-1707]{Martin Pohl}
\affiliation{University of Potsdam, Institute of Physics and Astronomy, D-14476 Potsdam, Germany}
\affiliation{Deutsches Elektronen-Synchrotron DESY, Platanenallee 6, 15738 Zeuthen, Germany}

\author[0000-0001-8867-2693]{Oscar Macias}
\affiliation{GRAPPA $-$ Gravitational and Astroparticle Physics Amsterdam, University of Amsterdam, Science Park 904, 1098 XH Amsterdam, The Netherlands}
\affiliation{Institute for Theoretical Physics Amsterdam and Delta Institute for Theoretical Physics, University of Amsterdam, Science Park 904, 1098 XH Amsterdam, The Netherlands}

\author{Phaedra Coleman}
\author[0000-0003-4864-5150]{Chris Gordon}
\affiliation{School of Physical and Chemical Sciences, University of Canterbury, Christchurch, New Zealand}

\begin{abstract}
We present a new reconstruction of the distribution of atomic hydrogen in the inner Galaxy that is based on explicit radiation-transport modelling of line and continuum emission and a gas-flow model in the barred Galaxy that provides distance resolution for lines of sight toward the Galactic Center. The main benefits of the new gas model are, a), the ability to reproduce the negative line signals seen with the H$I$4PI survey and, b), the accounting for gas that primarily manifests itself through absorption.

We apply the new model of Galactic atomic hydrogen to an analysis of the diffuse gamma-ray emission from the inner Galaxy, for which an excess at a few GeV was reported that may be related to dark matter. We find with high significance an improved fit to the diffuse gamma-ray emission observed with the \textit{Fermi}-LAT, if our new H$I$ model is used to estimate the cosmic-ray induced diffuse gamma-ray emission. The fit still requires a nuclear bulge at high significance. Once this is included there is no evidence for a dark-matter signal, be it cuspy or cored. But an additional
so-called boxy bulge is still favoured by the data.
This finding is robust under the  variation of various parameters, for example the excitation temperature of atomic hydrogen, and a number of tests for systematic issues.



\end{abstract}


\section{Introduction} \label{sec:intro}
Since its discovery some ten years ago \citep{2009arXiv0910.2998G,2011PhLB..697..412H}, the excess of gamma rays observed with the Fermi-LAT from the Galactic-Center region has remained one of the most intriguing open questions in astroparticle physics. Although published interpretations concentrate on a dark-matter interpretation or a millisecond pulsar related origin, there is no consensus on the origin of this so-called Galactic-Center excess (GCE). {See for example section 6 of \cite{Slatyer:2021qgc} for a review. } 
One of the main systematic difficulties is the need to accurately model the intense diffuse gamma-ray emission and the gamma-ray sources in the region. 

Several studies have claimed to find a non-Poissonian component to the GCE \citep{Bartels:2015aea,Lee:2015fea}, which
may be further evidence for the millisecond-pulsar explanation. However, there is some controversy regarding the level of systematics in this approach \citep[e.g.,][]{LeaneSlatyer2019,LeaneSlatyer2020,
Leane2020,Buschmann:2020adf,Chang2020,Calore:2021jvg, List:2021aer,Mishra-Sharma:2021oxe}.

{Gamma rays can be produced by cosmic-ray electrons and ions in what is referred to as leptonic and hadronic radiation processes. The main leptonic emission processes are inverse-Compton scattering of very high energy electrons off ambient photons and nonthermal bremsstrahlung \citep{1970RvMP...42..237B}. Hadronic emission processes involve the production of secondary particles in collisions of cosmic rays with gas nuclei and their eventual decay to gamma rays, which can be well modelled with Monte-Carlo event generators \citep[e.g.][]{2020APh...12302490B}. Both nonthermal bremsstrahlung and hadronic emission scale with the gas density, and so they provide the dominant contribution to the diffuse Galactic gamma-ray intensity for lines of sight through the Galactic plane and in particular toward the Galactic Center region, where the gas column density is very high.}
Modelling the diffuse interstellar gamma-ray emission thus requires knowledge of the distribution of gas in the Galaxy, which must be convolved with the spatial distribution of cosmic rays to estimate the gamma-ray emissivity along each line of sight. {Ionized gas is seen in the dispersion of the radio signals from pulsars. Line spectra of atomic hydrogen (H$I$) or CO as tracer of molecular hydrogen provide information on the line-of-sight velocity of the gas, whereas what is needed is the distribution along the line of sight. Traditionally the Doppler shift of the line signal is modelled assuming circular motion around the Galactic Center \citep[e.g.][]{2006PASJ...58..847N}, which fails toward the inner Galaxy, on account of the vanishing line-of-sight component of the flow velocity.}
\citet{2008ApJ...677..283P} used a {model of noncircular} gas flow based on the smoothed particle hydrodynamic (SPH) simulations described in \citet{2003MNRAS.340..949B} to deconvolve CO data. They employed an iterative method to successively reduce signal in the line spectrum and place it at the eight best-matching distance intervals, until there is only noise left. In \citet{2018NatAs...2..387M} an analogous deconvolution of H$I$ data was found to provide a better fit to the diffuse gamma-ray emission from the Galactic-Center region than do the gas maps of the standard \textit{Fermi}-LAT data analysis pipeline\footnote{\url{https://fermi.gsfc.nasa.gov/ssc/data/access/lat/BackgroundModels.html}} \citep[see also][]{2018ApJ...856...45J}. 

{The available line spectra are essentially spectral distributions of observed intensity minus the wide-band continuum emission that may stem from synchrotron radiation or thermal bremsstrahlung. Each radiation process can provide emission and absorption, and the observed intensity reflects the balance of all emission and all absorption processes along the line of sight. In the earlier analysis \citep{2018NatAs...2..387M},} the absorption correction {for the line signal} was minimal and involved only self-absorption with constant excitation temperature $T_\mathrm{exc}=170\,$K. Continuum emission was ignored, which means a weak positive signal was deemed optically thin and a negative signal had to be disregarded. In the Galactic-Center region these simplifications lead to a potentially significant underestimation of the mass of atomic gas, and hence a deficit in the predicted diffuse gamma-ray emission and an artificial indication for new emission components. 

In this paper we present an advanced model of atomic gas in the Galaxy and apply it to the analysis of gamma-ray emission from the Galactic Center. We account for both line and continuum emission in the radiation transport, which allows the modelling of negative line intensity and traces gas in both emission and absorption. For better comparison with the results of \citet{2018NatAs...2..387M} we retained with minor modifications the gas-flow model and rotation curve as used in \citet{2008ApJ...677..283P}, although newer studies of gas flow and Galactic rotation had been published \citep[e.g.][]{2010PASJ...62.1413B,2014MNRAS.444..919P,2015PASJ...67...75S,2020arXiv201215770M}. The recently observed radial flow beyond the solar circle \citep{2020A&A...642A..95C} should not be relevant for the lines of sight toward the inner Galaxy that we consider here. 

\section{Method}
\subsection{Radiation transport}
{In the absence of scattering, the evolution of the intensity, $I$, along a line of sight, $s$, can be described by the transport equation}
\begin{equation}
\frac{dI}{ds}=j_c+j_l -\alpha_l I\ ,
\label{eq:radtrans}
\end{equation}
where we allow for continuum emission with coefficient $j_c$, line emission, $j_l$, and absorption, $\alpha_l$.
{Given the observed line intensity spectrum, we want to infer the density of atomic hydrogen along the line of sight, $n_{\mathrm{H}I}(s)$, because that is an important scaling factor for the gamma-ray emissivity. The emission coefficient, $j_l(s)$, is proportional to $n_{\mathrm{H}I}(s)$. All quantities in eq.~\ref{eq:radtrans} are functions of the radiation frequency, $\nu$, and so the radiation transport equation must be solved for each frequency. The gas-flow model provides the distances from which the H$I$ line signal may arrive with the appropriate Doppler shift to appear at the frequency in question, and so both the gas flow and the particulars of radiation transport must be understood to transform the observed line spectra into the line-of-sight density distribution of atomic gas. In earlier models of diffuse Galactic gamma-ray emission, including our own, continuum emission was typically ignored, $j_c=0$. That is clearly a gross simplification, because intense continuum emission is observed toward the inner Galaxy. }

{For a thermal gas the ratio of emission and absorption coefficients is a Planckian with the excitation temperature, $T_\mathrm{exc}$ (see for example section 7.4.1 of \citet{Draine2011}). We find good fits to the observed H$I$ line spectra for a broad range of excitation temperatures, $T_\mathrm{exc}\gtrsim 180\,$K, as did earlier studies of compact H$I$ absorption features in the spectra \citep{2005ApJ...626..214G}. For continuum emission the excitation temperature is much higher than that, for example $10^4\,$K for free-free emission and millions of degrees for synchrotron radiation. The continuum absorption coefficient then is very small, $\alpha_c\simeq 0$, and in eq.~\ref{eq:radtrans} we ignore it entirely.} 
In our model the line of sight is binned, and the radiation coefficients are assumed to be constant within a bin. At the front of each bin of length $\Delta s$, corresponding to an optical depth $\tau=\Delta s\, \alpha_l$, we find for $\tau > 0$
\begin{equation}
I(\Delta s)=I_0 \exp(-\tau) + \frac{j_c+j_l}{\alpha_l}\,\left[1-\exp(-\tau)\right]\ ,
\end{equation}
where $I_0$ is the intensity at the rear boundary.


The change of the intensity along the bin is given by
\begin{equation}
\begin{aligned}
\Delta I &=I(\Delta s)-I_{0} \\
&=\left(\frac{j_{c}+j_{l}}{\alpha_{l}}-I_{0}\right)\left[1-\exp(-\tau)\right] \\
&=
\left(\frac{j_c \Delta s+j_l \Delta s}{\tau}-I_{0}\right)\left[1-\exp(-\tau)\right]
\end{aligned}
\end{equation}

We are working in the low-frequency limit where the brightness temperature $T_B$ is proportional to the intensity {and is commonly used as a proxy for it. This is convenient, because the brightness temperature directly relates to the excitation temperature, $T_\mathrm{exc}$.} The increment in $T_B$ then is
\begin{equation}
\Delta T_B= \left(\frac{\Delta T_c+\Delta T_l}{\tau}-T_0\right)\,\left[1-\exp(-\tau)\right]\ ,
\label{eq:radtransTB}
\end{equation}
where  $T_0$ is the brightness temperature at the rear boundary of the bin and $\Delta T_c\propto j_c\,\Delta s$ is the increment in continuum brightness temperature along the bin. The increment in the line brightness temperature is given by $\Delta T_l = \tau\,T_\mathrm{exc}$, where $T_\mathrm{exc}\propto j_l  / \alpha_l$ is the excitation temperature of the atomic-hydrogen gas.

For each velocity bin and line of sight, we successively apply eq.~\ref{eq:radtransTB} to find the brightness temperature at Earth, from which we subtract the continuum temperature to obtain line spectra,
\begin{equation}
T_l=-T_c+\sum_i \Delta T_B(s_i) \ ,\quad T_c=\sum_i \Delta T_c(s_i)\ ,
\label{eqtb}
\end{equation}
where $T_l$ is the observed line brightness temperature.
Matching those spectra to the observed line spectra will yield $\Delta T_l$ for each distance bin. The corresponding contribution to the column density of gas, $N_H$, is
\begin{equation}
\Delta N_H=(1.8\cdot 10^{18}\ \mathrm{s\,K^{-1}\,cm^{-2}\,km^{-1}})\,\Delta v\,\Delta T_l \ ,
\end{equation}
where $\Delta v$ is the bin width in velocity space \citep{1990ARA&A..28..215D}. {Correlations between the gas distribution along neighboring lines of sight likely exist \citep{mertsch2022bayesian}, but are not considered here to retain the structural simplicity of the deconvolution process.}

We use data from the H$I$4PI survey \citep{2016A&A...594A.116H}, that {essentially is a merged dataset composed of the Effelsberg-Bonn H$I$ Survey \citep[EBHIS,][]{2011AN....332..637K,2016A&A...585A..41W} and the Galactic All-Sky Survey \citep[GASS,][]{2009ApJS..181..398M,2015A&A...578A..78K}. The observations were made with the Effelsberg 100-m telescope and the Parkes 64-m telescope. Using large single-dish telescopes has the benefit of providing a high sensitivity, a decent angular resolution, and excellent coverage of large-scale emission features. Like the Leiden-Argentina-Bonn survey (LAB) that is supersedes, the H$I$4PI survey is corrected for stray-radiation \citep{1980A&A....82..275K}, using the same methodology.}
The H$I$4PI survey
outperforms the LAB survey in angular resolution ($16.2\arcmin$~FWHM), sensitivity ($\sigma_\mathrm{rms}\simeq 43$~mK), and by its full spatial sampling. Our data cube has an angular sampling of $\Delta l=\Delta b = 5\arcmin$ and a velocity binning of $\Delta v\simeq 1.3$~km/s.

\subsection{Continuum Modelling}
The continuum modelling was performed as follows:
We have
the CHIPASS\footnote{\url{https://lambda.gsfc.nasa.gov/product/foreground/fg_chipass_info.cfm}} \citep{2014PASA...31....7C} and Stockert\footnote{\url{https://lambda.gsfc.nasa.gov/product/foreground/fg_stockert_villa_info.cfm}} \citep{2001A&A...368.1123T,2001A&A...376..861R} continuum datasets. {The CHIPASS 1.4 GHz continuum map covers the sky south of +25 degrees declination with a resolution of 14.4 arcmin and a sensitivity of about 40 mK. The map results from a reprocessing of archival data from the H$I$ Parkes All-Sky Survey (HIPASS) and the H$I$ Zone of Avoidance (HIZOA) survey. The Stockert survey combines observations of the northern sky with the 25-m Stockert telescope and the southern sky with the 30-m Villa-Elisa telescope. The effective resolution is about 35 arcmin, the effective sensitivity is about 50 mK, and the zero level accuracy is 0.5 K.}

CHIPASS data were reprojected to match the Stockert data. {The original CHIPASS map is blanked for regions that were not observed, for example near the peaks of 16 strong compact sources where the signals were saturated. These sources include Sgr A$^\ast$.
We patched} missing regions in the CHIPASS data with the Stockert data using a $\tanh$ based modulating function
to smoothly combine the two data sets.
We then  fitted a model of the continuum emission consisting of three Gaussian disk components,
\begin{equation}
  \Delta T_c/\Delta s= 
    \sum_{i=1}^3 a_i \exp \left[-\frac{1}{2} \left(\frac{r^2}{\sigma _{r,i}^2}+\frac{z^2}{\sigma_{z,i}^2}\right)\right]    
    \label{eq:ThreeDisks}
\end{equation}
where $a_i$, $\sigma_{r,i}$, and $\sigma_{z,i}$ were the fitted model parameters. The model was fitted in Galaxy-centered cylindrical coordinates with radius, $r$, and height above the Galactic plane, $z$. Using least-squares optimization 
on a ROI of $|l|<50^\circ$, $|b|<25^\circ$, we fitted one disk, then fixed that and fitted a second one. 
We then fixed those two components and fitted a third component initialized to the extent of the high-intensity region in the Galactic Center.
Then we freed all parameters to fit the three components simultaneously. The best fit parameters are shown in Table~\ref{Tab:ThreeDisks}.

\begin{figure}
    \centering
    \begin{tabular}{c}
    
    \includegraphics[width=0.95\linewidth]{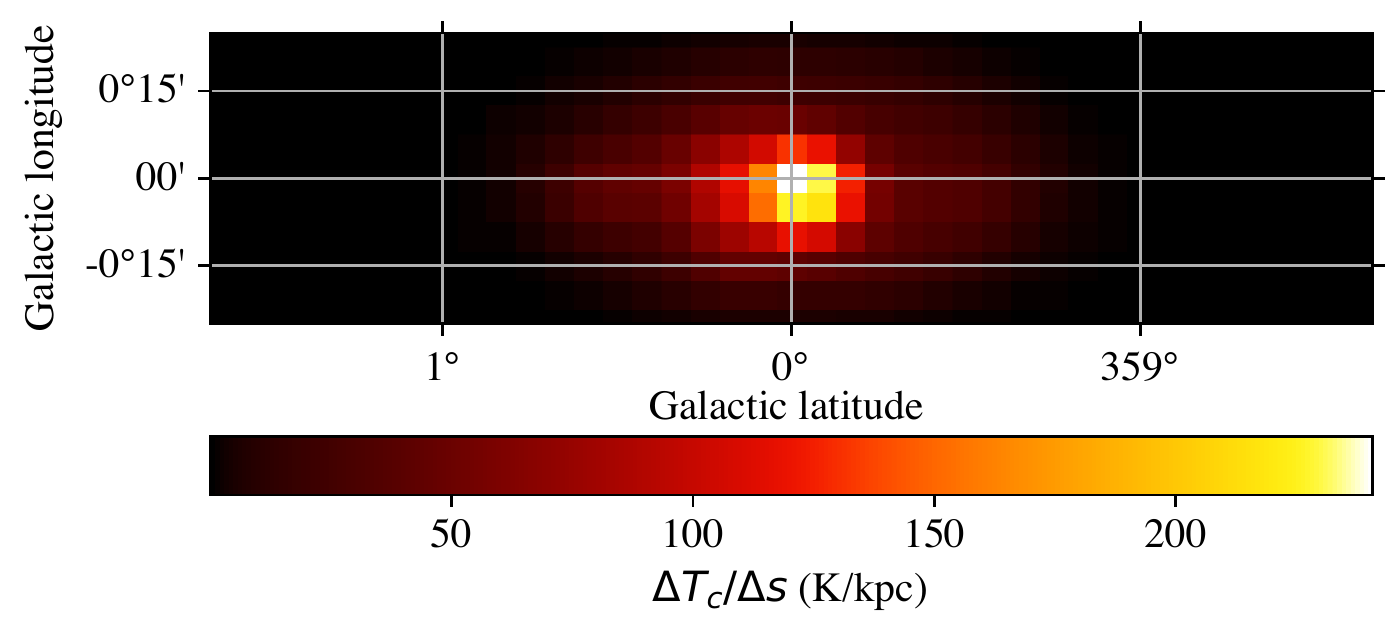}\\  \\
    \includegraphics[width=0.95\linewidth]{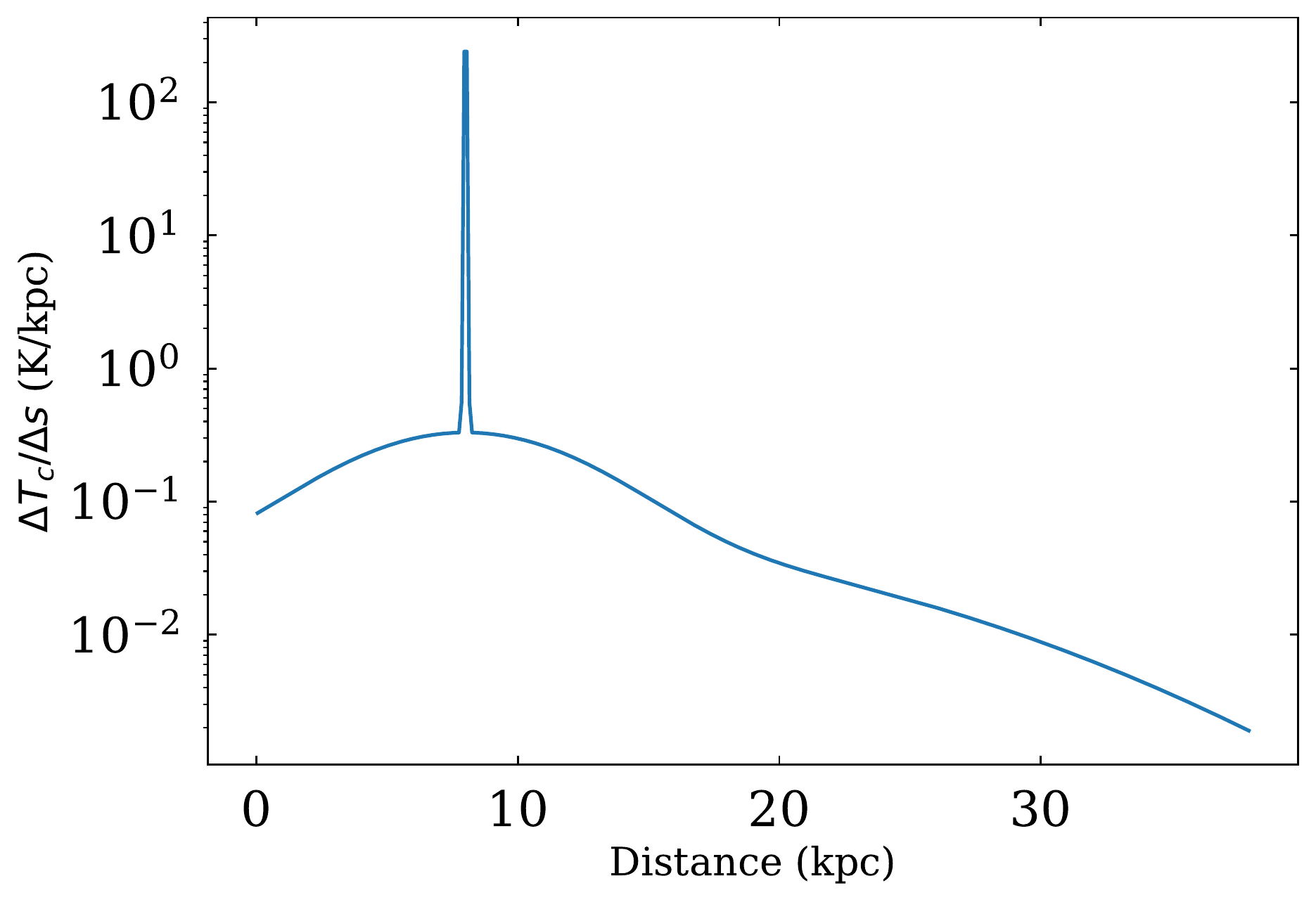}\\
    
    \end{tabular}
    \caption{Continuum emission model. Top: Cross section of $\Delta T_c/\Delta s$ at a distance of 8 kpc from the solar system. 
    Bottom: Profile of $\Delta T_c/\Delta s$ for $l=0^\circ$ and $b=0^\circ$. 
    }
    \label{fig:cont}
\end{figure}

\begin{deluxetable*}{cccc}
\tablecaption{{Best fit parameter values for the three-disk continuum model. \label{Tab:ThreeDisks}}}
\tablehead{
\colhead{Parameter}   & \colhead{Amplitude ($a$)} &  \colhead{Radial standard deviation ($\sigma_r$)}  &  \colhead{Vertical standard deviation ($\sigma_z$)}  
 \\
\colhead{Units} & \colhead{K/kpc} &\colhead{kpc} &\colhead{kpc} }
\startdata
Disk 1 & 0.29& 12& 5.2\\
Disk 2& 3100& 0.038&   0.021\\
Disk 3& 1.5 & 4.0& 0.13   \\
\enddata
\tablecomments{
The $a$ parameters in Eq.~\ref{eq:ThreeDisks} were fitted so that after line-of-sight integration in any direction one obtains the brightness temperature. Hence the units of $a$ are K/kpc.}
\end{deluxetable*}
Using this three-Gaussians model, we created an instance of the model on the grid of the gas-deconvolution cube and renormalized it to the observed continuum temperature for each line of sight, and so the signal in the bins sums up to the observed continuum brightness. As the model is just a simple three-component model, the continuum cube has a few stark point-source-like components when renormalized, which manifest themselves as bright streaks through slices of the cube. The Galactic Center is a hotspot in continuum brightness with $T_c \gtrsim 500\,$K.  Figure~\ref{fig:cont} displays the distribution of continuum emissivity per distance bin, which likewise has a sharp peak at the Galactic Center. 

\subsection{Algorithm}
For each bin in velocity space in which the modulus of the signal exceeds $0.15$~K, we find the eight best-fitting distance solutions in distance bins of 50~pc. The signal is then distributed over those distance solutions using weights that are calculated as in \citet{2018NatAs...2..387M}. 

Tests show that the nonlinearity in the radiation transport very much complicates accounting for a finite width of the signal from individual clouds and correlations between neighboring lines of sight, at least compared to a Bayesian inference of CO line data \citep{2020arXiv201215770M}. To avoid artefacts and a strong dependence on priors we treat each line of sight and velocity bin independently. As we ignore proper motion of gas clouds relative to the local average flow, there is more signal without a distance solution than with the deconvolution technique of \citet{2008ApJ...677..283P}. This signal is placed according to the distance solutions at the closest velocity covered in the gas-flow model, but the radiation transport is separately calculated for each velocity bin. 
Then the distance resolution is reduced by a factor two by combining neighboring distance bins. This fixes the line-of-sight distribution of the signal.

\begin{figure}
    \centering
    \includegraphics[width=0.98\linewidth]{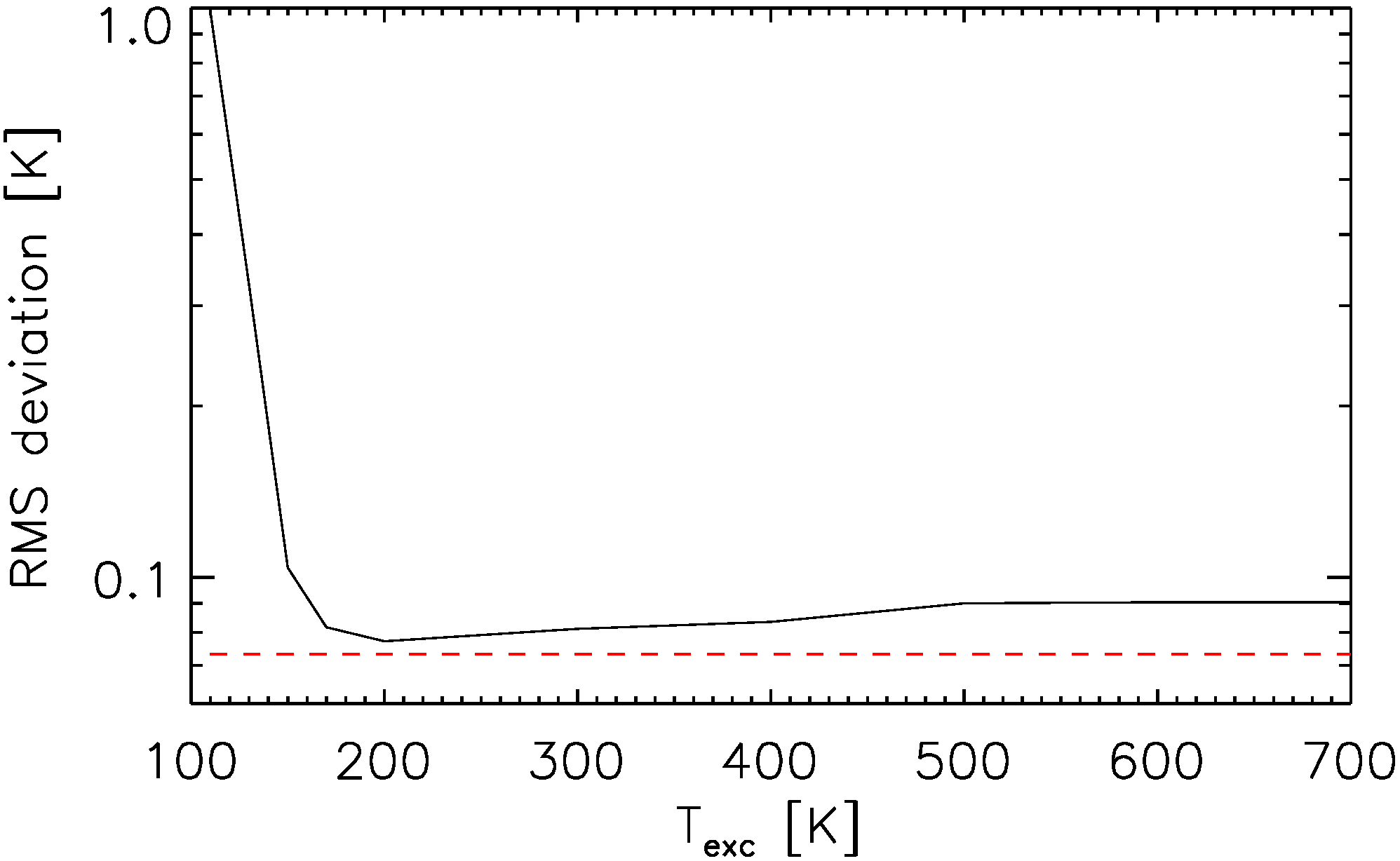}
    \caption{Averaged difference between the model spectra and the observed spectra, plotted as a function of excitation temperature, $T_\mathrm{exc}$. The red dashed line indicates the mismatch level for
an excitation temperature that is allowed to vary with longitude, $l$ and latitude, $b$.}
    \label{fig:noise}
\end{figure}

To determine the amplitude of the signal we create a set of 60 logarithmically spaced model signals with integrated line emissivity, $\smallint ds\,j_l$, ranging from $0.15$~K to more than $10^3$~K, for which we solve eq.~\ref{eqtb}.
We then search for the one model temperature, $T_l$, that best matches the observed brightness as given in the H$I$4PI data. If the observed value is beyond the range of model values, we pick the closest one, otherwise we use linear interpolation between the two nearest models.
The rms deviation between model and observed spectrum is computed and retained as accuracy parameter to the model file. In Fig.~\ref{fig:noise} we display the mean mismatch as a function of the excitation temperature, $T_\mathrm{exc}$. Averaged over the area of interest, we see the smallest deviation with $T_\mathrm{exc}=200\,\mathrm{K}$, and somewhat higher temperatures work nearly as well.

Note that we place gas seen in absorption also on the far side, although it does not provide much absorption there because it is likely behind the continuum source. Otherwise we would construct a near-heavy Galaxy in regions of high absorption. 

\begin{figure}
    \centering
    \includegraphics[width=0.99\linewidth]{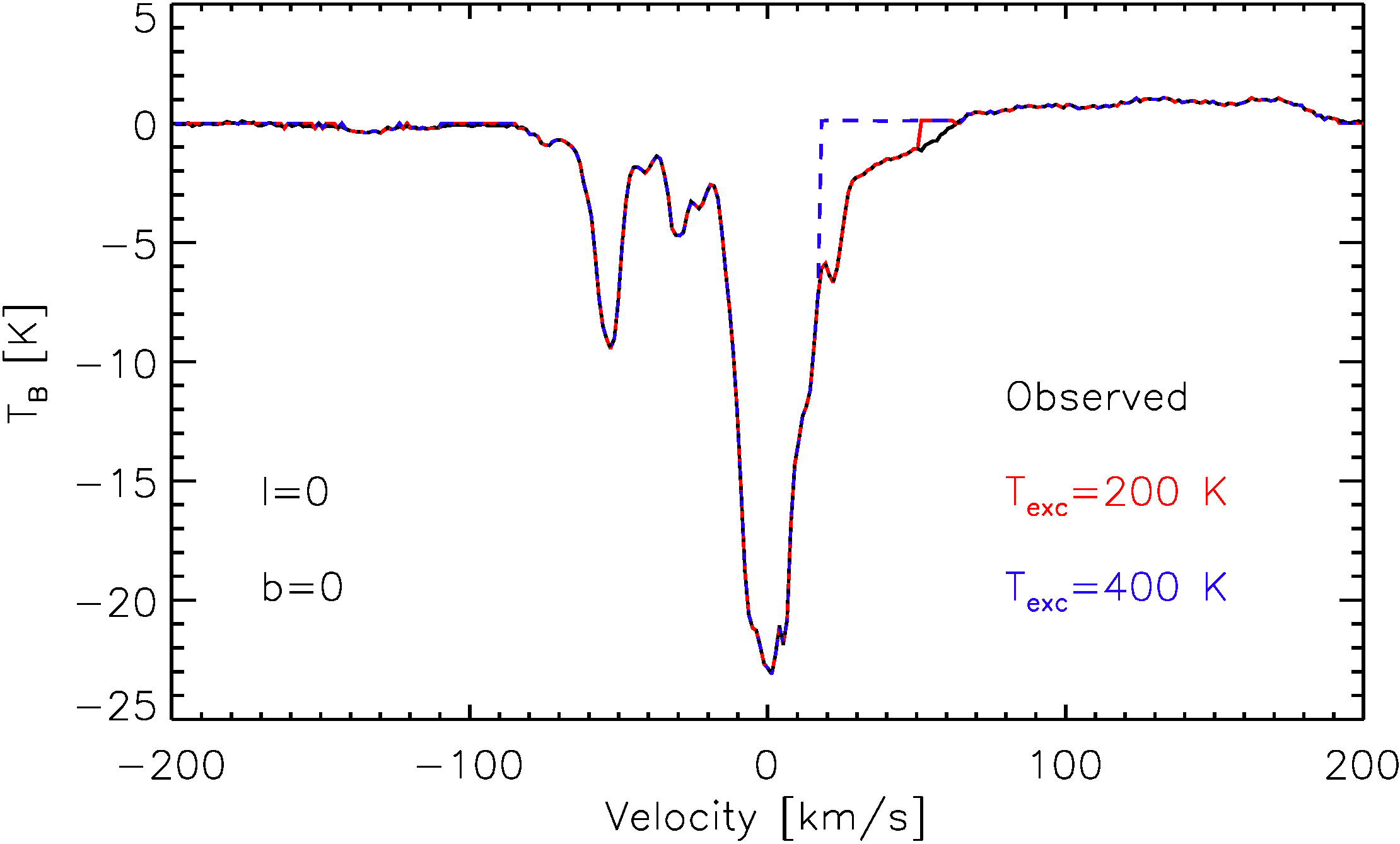}
     \includegraphics[width=0.99\linewidth]{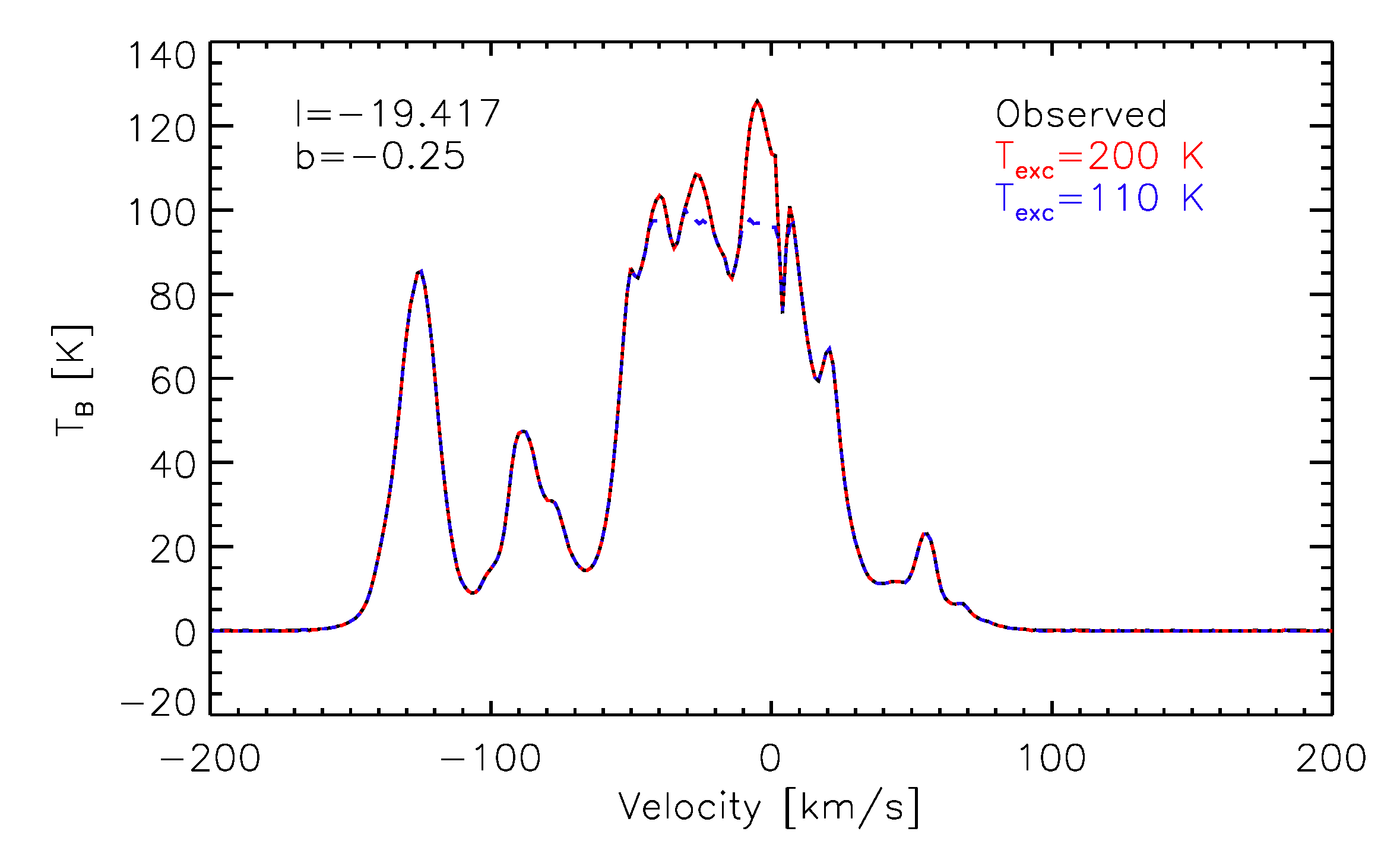}
    \caption{Comparison of the observed and the modelled H$I$ spectrum toward the Galactic Center (top panel)
    and a line of sight with a high-intensity peak (bottom panel). }
    \label{fig:profiles}
\end{figure}

Figure~\ref{fig:profiles} presents for two values of the excitation temperature, $T_\mathrm{exc}$, the modelled and observed H$I$ spectra for the line of sight toward the Galactic Center and a second area with high line intensity. To be noted from the top panel of the figure is that solving the radiation transport equation with continuum emission can reproduce strong absorption features and provide an estimate of the H$I$ column density where absorption occurs. Modelling negative line intensity becomes difficult for higher values of $T_\mathrm{exc}$, in particular for velocities for which the distance solutions are predominantly behind the region of high continuum emissivity. Whereas for $T_\mathrm{exc}=200\,$K that happens in only a narrow band around $v=50\,\mathrm{km\,s^{-1}}$, implying that only a small fraction of the gas is poorly modelled, a larger mismatch is 
seen for $T_\mathrm{exc}=400\,$K.

The signal at $(l,b)=(0^\circ,0^\circ)$  in the top panel is almost perfectly fit by $T_\mathrm{exc}=110$K but that excitation temperature doesn't work well for other lines of sight with high intensity peaks, an instance of which is shown in the bottom panel.

In reality one should expect to find gas clouds on the line of sight that have different excitation temperatures. It is quite conceivable that the absorption feature around $v=50\,\mathrm{km\,s^{-1}}$ 
in the top panel of figure~\ref{fig:profiles}
is caused by relatively little cold gas immediately in front of the Galactic Center, whereas most of the gas clouds have temperatures of a few hundred Kelvin. {Accounting for variations in the excitation temperature along the line of sight would introduce a large number of free parameters that are, if at all, only constrained by the quality of the reconstruction of the H$I$ line spectra. Tests involving a thin layer of cold ($50$~K) gas in the foreground or alternating $T_\mathrm{exc}=100$K and $T_\mathrm{exc}=200$K every 200 pc along the line of sight gave extremely poor reconstructions of the line spectra, suggesting that improvements in the spectral reconstruction may be achieved only in very few compact regions. Compact clouds of atomic hydrogen can be detected as small-scale self-absorption features in H$I$ surveys with very high resolution \citep{2005ApJ...626..214G}. Faint self-absorption features are detected almost everywhere, provided the background is bright enough, whereas strong features ($\Delta T > 20$~K) are few and possibly associated with molecular condensation in spiral arms \citep{2005ApJ...626..195G}. The paucity of strong absorption features is in line with the widespread abundance of line signal stronger than $80$~K, that significant amounts of colder gas in the foreground wouldn't allow.  Corrections for hydrogen self-absorption have been made in an analysis of gamma-ray emission from RX~J1713.7-3946 at $l\approx 347^\circ$, and the effect was found to be minuscule with less than 0.5\% difference in the fit residuals \citep[see Fig.2 in][]{2015A&A...577A..12F}. The above suggests it is reasonable to use a uniform excitation temperature for a particular line of sight. The effect of variation in the excitation temperature are likely moderate, of the same order as that of the choice of excitation temperature itself. In the appendix we show that a constant $T_\mathrm{exc}=200$K, i.e. not varying with $l$ and $b$, leads to quantitatively comparable results for the statistical significance of the GCE templates.}

\begin{figure}
    \centering
    \includegraphics[width=0.99\linewidth]{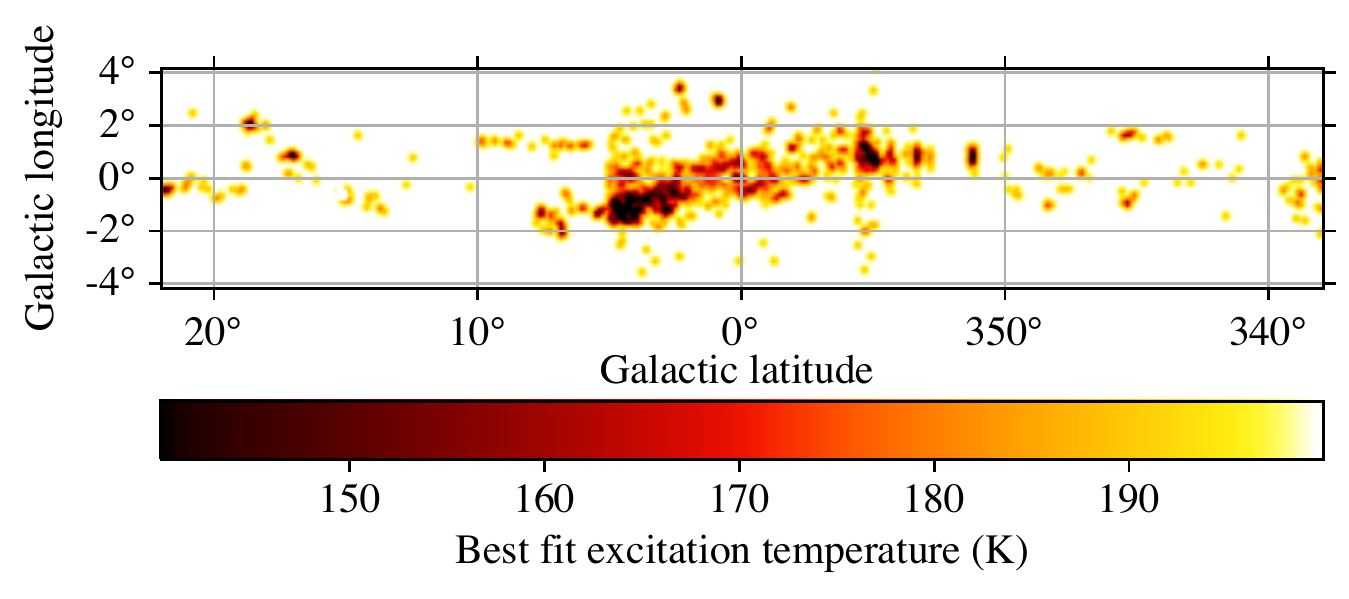}
    \caption{Best-fit excitation temperature for each line of sight. The image was smoothed with a 0.08$^\circ$ Gaussian filter and cropped to contain 99.5\% of intensity values for display purposes.}
    \label{fig:best_Texc}
\end{figure}

In Fig.~\ref{fig:best_Texc} we show a map of the best-fit $T_\mathrm{exc}$ as a function of $l$ and $b$. Only $|b|\leq 4^\circ$ is shown, as higher latitude areas in our region of interest almost all had $T_\mathrm{exc}=200K$.

\subsection{Fermi-LAT analysis}
{We used eight years (August 4, 2008$-$August 2, 2016) of \textsc{Pass 8 Release 3 (P8R3) ULTRACLEANVETO}-class photon data in the energy range 667 MeV$-$158 GeV. Photons detected at zenith angles $>90^{\circ}$ were removed to reduce the contamination from gamma-rays generated by cosmic ray interactions in the Earth's atmosphere. Additionally, we applied the recommended data filters (DATA$_{-}$QUAL$>$0)\&\&(LAT$_{-}$CONFIG==1).}

{The data reduction as well as the data analysis were performed with the \textsc{Fermitools v1.0.1}\footnote{\url{https://github.com/fermi-lat/Fermitools-conda/wiki}} package and instrument response functions \textsc{P8R3$_{-}$ULTRACLEANVETO$_{-}$V2}. The ROI of the analysis is defined by a square region of size $40^{\circ}\times 40^{\circ}$ centered at Galactic coordinates $(l,b) = (0,0)$. We used a binned-likelihood method  with a spatial binning of $0.2^\circ$ and 15 logarithmically-spaced energy bins (in the range 667 MeV$-$158 GeV, corresponding to $\Delta E/E\simeq 0.44$). Details about the statistical procedure and all other astrophysical templates considered are given in the Appendix.\footnote{{The analysis templates are publicly available at \url{https://doi.org/10.5281/zenodo.6276721}.}} }    

\section{Results}

\begin{figure}
    \centering
    \includegraphics[width=0.96\linewidth]{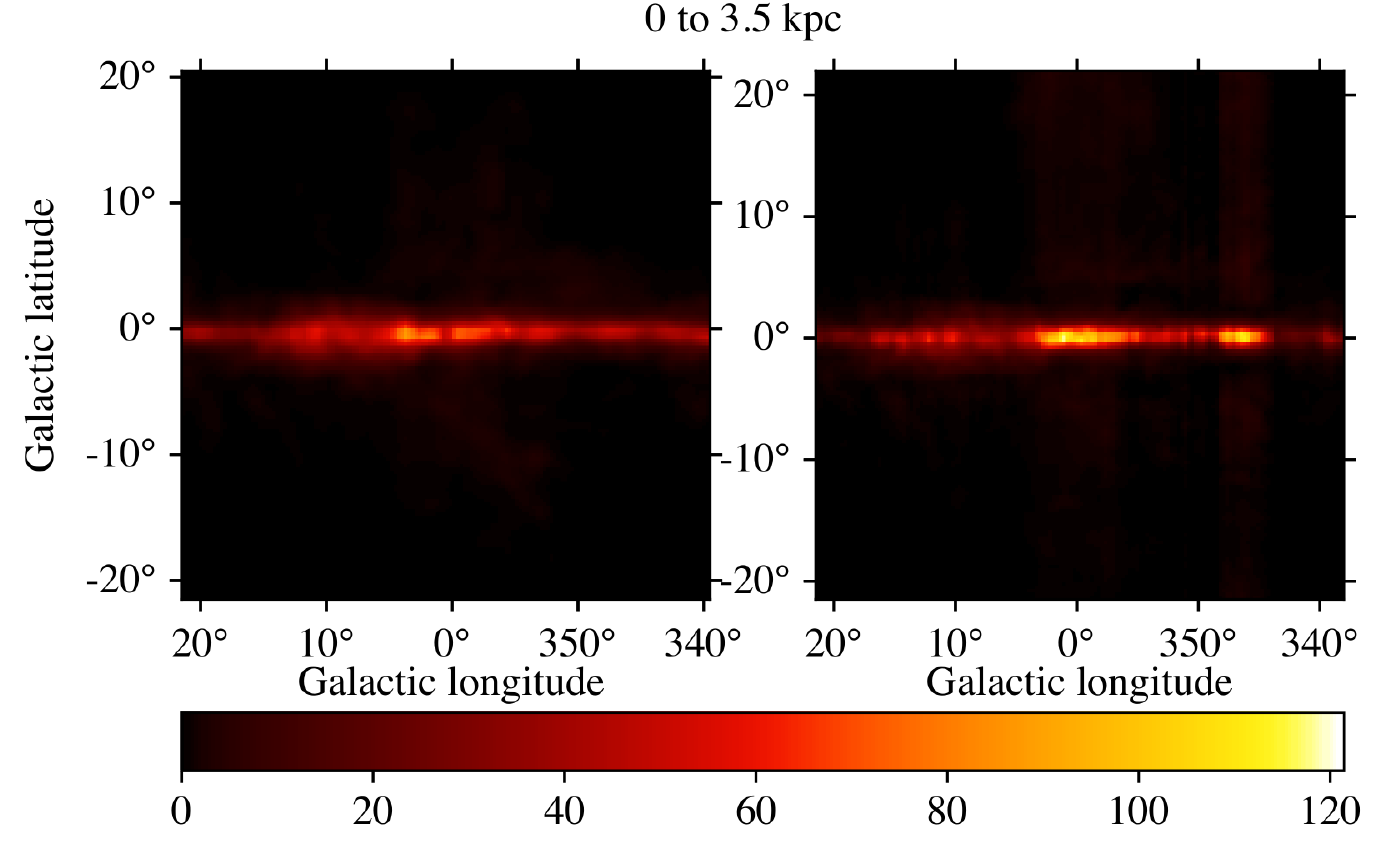}
       \includegraphics[width=0.96\linewidth]{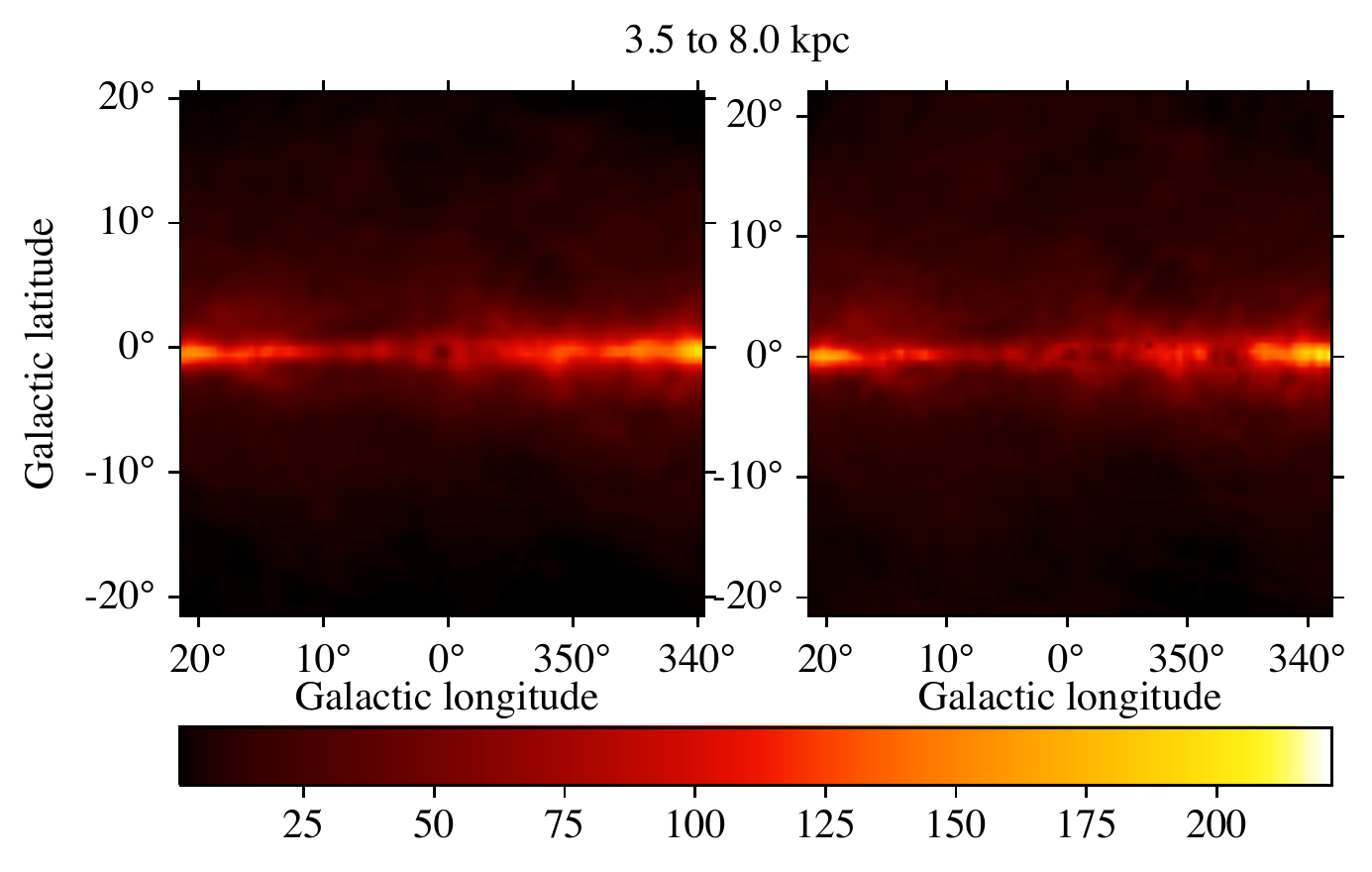}
        \includegraphics[width=0.96\linewidth]{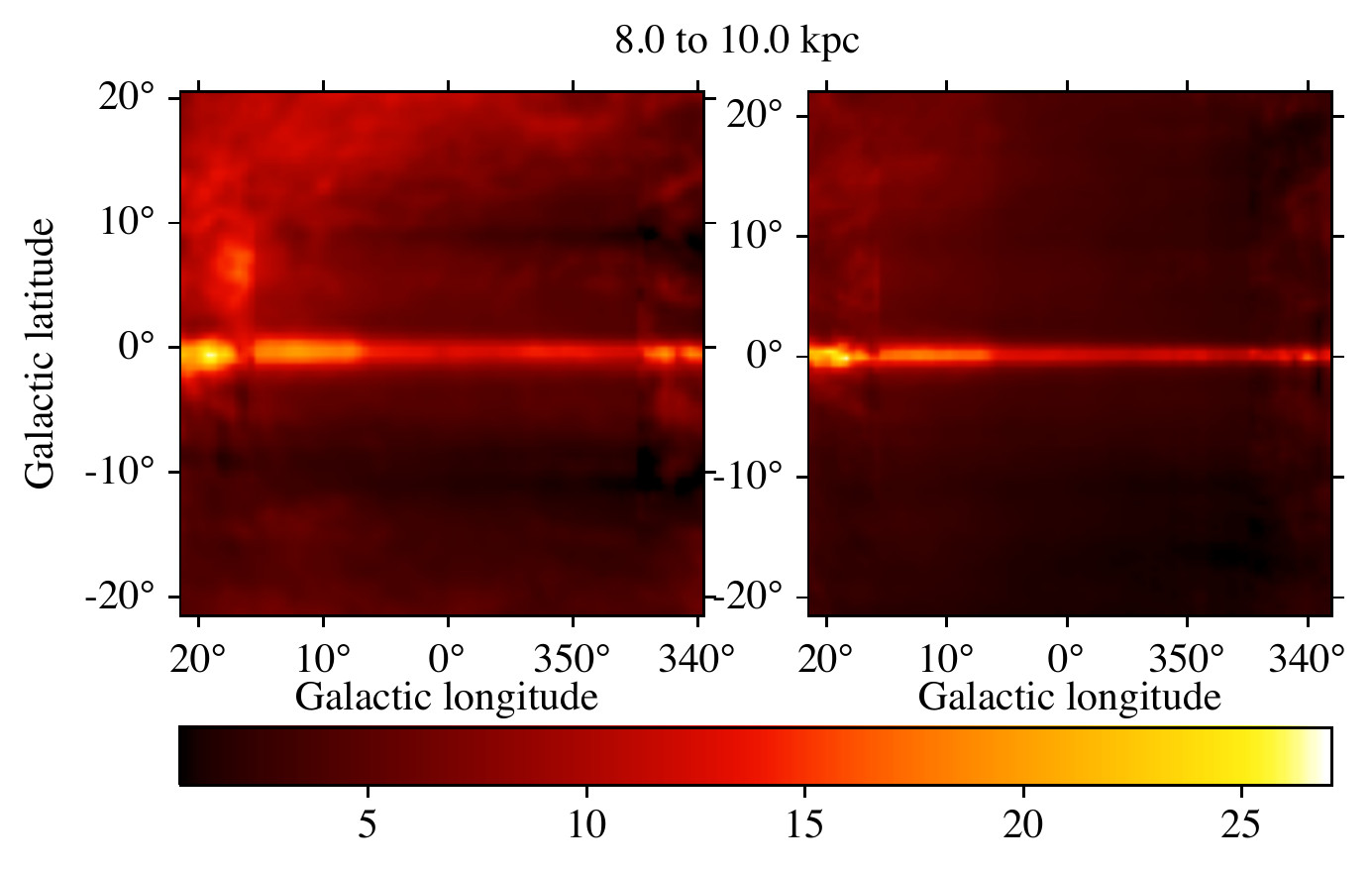}
       \includegraphics[width=0.96\linewidth]{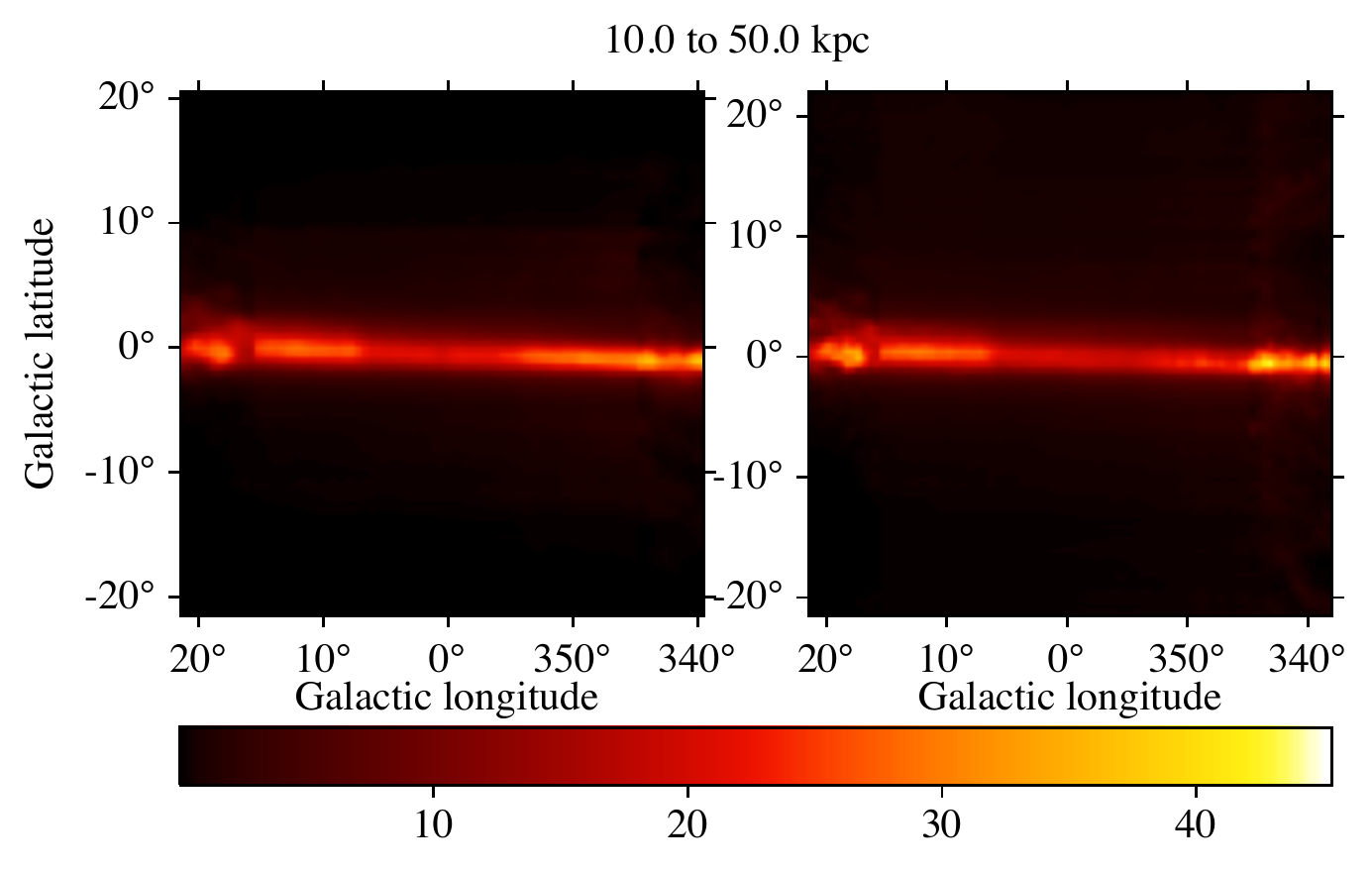}
    \caption{{Maps of column density in units of $10^{20}$cm$^{-2}$ for four concentric rings. The right panels show models that account for continuum emission with $T_{\rm exc}$ allowed to vary with $l$ and $b$. The left panels display models from \cite{2018NatAs...2..387M} that do not account for continuum emission.}}
    \label{fig:coldens}
\end{figure}

In the previous sections, we presented our improved model for the distribution of H$I$ in the Galaxy. The new templates trace gas in both emission and absorption, and account for negative line intensity. Additionally, we have upgraded the dust correction maps with the use of \textsc{planck} data, but we emphasize that the molecular hydrogen maps in this work are still the same as those in~\cite{2018NatAs...2..387M}. Likewise, the fit parameters for the cosmic-ray spectra, and hence the gamma-ray emissivity are still considered constant in concentric rings of galactocentric radius. For comparison, we display in Fig.~\ref{fig:coldens} as a function of $l$ and $b$ the H$I$ column density that is attributed to four of these concentric rings. The new model, shown on the right, has considerably fewer artefacts than has the older model. {The artefact at $l\simeq 15^\circ$ and $b\simeq -0.6^\circ$ arises from a compact source of high continuum flux in this direction whose position on the line of sight is likely not well represented by the azimuthal symmetry in the continuum emissivity model.} 
To be noted is the enhanced column density attributed to the Galactic plane at $r\le 3.5$~kpc, where H$I$ absorption is particularly strong. Within a few degrees from the Galactic Center, this signal is not simply taken from any of the other concentric rings. Instead it results from the proper modelling of H$I$ absorption and the strong continuum emission from that direction\footnote{{The deconvolved H$I$ data cube is publicly available at \url{https://doi.org/10.5281/zenodo.5845040}.}}.

In this section, we perform fits to gamma-ray data from the Galactic Center region in order to evaluate the impact of the new templates on the characteristics of the GCE. 

\begin{table}
\centering
\begin{tabular}{cccc}
\hline\hline
Baseline  & Additional &  $\Delta \mathrm{TS}$  &  Significance \\ 
model               &   source                 &               &   \\\hline
Base  & Cored ellips. & 0.0              & $0.0\;\sigma$\\
Base  & Cored             & 0.1              & $0.0\;\sigma$\\
Base  & BB                &  282.2             & $15.3\;\sigma$\\ 
Base  & NFW ellips.   &  647.2             & $24.2\;\sigma$\\ 
Base  & NFW               &  807.1             & $27.3\;\sigma$\\ 
Base  & NB                &  1728.9             & $40.8\;\sigma$\\ \hline
Base+NB  & Cored ellips. &    0.1           & $0.0\;\sigma$\\ 
Base+NB  & Cored             &    0.7           & $0.0\;\sigma$\\
Base+NB  & NFW ellips.   &    1.0             & $0.0\;\sigma$\\ 
Base+NB  & NFW               &    3.4           & $0.2\;\sigma$\\ 
Base+NB  & BB                &    261.0           & $14.7\;\sigma$\\\hline 
Base+NB+BB  & NFW ellips.   &    0.1            & $0.0\;\sigma$\\ 
Base+NB+BB  & Cored ellips. &    0.4           & $0.0\;\sigma$\\ 
Base+NB+BB  & Cored             &    0.7           & $0.0\;\sigma$\\    
Base+NB+BB  & NFW               &    2.6           & $0.1\;\sigma$\\ 
\hline\hline
\end{tabular}
\caption{{\bf Statistical significance of the GCE templates for the H$I$ maps with varying $T_{\rm exc}$.} The Base model comprises the new hydrodynamic gas maps introduced in this work (divided in four concentric rings), dust correction maps, inverse Compton maps, the 4FGL point sources, and templates for the Fermi Bubbles, Sun, Moon, Loop I, and isotropic emission (see Appendix~\ref{sec:astrotemplates}). Additional sources considered in the analysis are: Nuclear bulge (NB)~\citep{Nishiyama2015}, boxy bulge (BB)~\citep{Coleman:2019kax},  NFW profile with $\gamma=1.2$, cored dark matter~\citep{Read:2015sta}, and ellipsoidal versions of these (see Fig.~3 in~\citep{Abazajian:2020tww}).
Note that as usual, all dark matter model templates are squared as is appropriate for pair-pair annihilation.
}\label{tab:loglike-values}
\end{table}

\subsection{Implications for the GCE}
\label{subsec:implicationsGCE}

To evaluate the impact of the new gas maps on the GCE, we include in our Base model the H$I$ maps that best reproduce the observed line emission, namely those with $T_{\rm exc}$ varying as a function of $l$ and $b$. Similar tests for the best-fitting constant excitation temperature ($T_{\rm exc}=200$ K) are presented in Appendix~\ref{sec:200K}. In addition, the Base model includes positive and negative dust correction templates~\citep{Fermi-LAT:4FGL},
3D inverse Compton (IC) maps divided in six concentric rings~\citep{Porter:2017vaa}, the Fourth Fermi Catalogue (4FGL) of point sources~\citep{Fermi-LAT:4FGL}, a Fermi Bubbles (FB) template~\citep{Macias:2019omb} based on that reconstructed in~\cite{Fermi-LAT:2014sfa}, specialized templates for the Sun and Moon, an isotropic emission model, and a geometrical template for Loop I (see Appendix~\ref{sec:astrotemplates}). 

We first start by running the bin-by-bin procedure, described in Appendix~\ref{sec:fermianalysis}, with the Base model. This is done by varying the flux normalization of all the point sources and extended templates such that the log-likelihood is independently maximized in each energy bin, using the \textit{Fermi} \textsc{pylikelihood} tool\footnote{\url{https://fermi.gsfc.nasa.gov/ssc/data/analysis/scitools/extended/extended.html}}. The next step consists of implementing the bin-by-bin method with an augmented model that includes the GCE templates. We consider four classes of dark matter (DM) profiles, and two maps tracing the distribution of stars in the inner Galaxy (all described in Appendix~\ref{sec:astrotemplates}). The statistical significance for each new source is obtained by computing the probability of $\Delta \mathrm{TS}$ as shown in Eq.~2.5 of \cite{Macias:2019omb}, and noting that each additional template has 15 degrees of freedom. We stress that when computing the statistical significance of each of the GCE templates, we simultaneously vary the fluxes of the baseline model and additional templates. We show the results of this step in the first six rows of Table~\ref{tab:loglike-values}, where the GCE templates are sorted according to their statistical significance. 

The fact that some of the additional templates are found with such a high statistical significance suggests that the Base model alone is insufficient to explain the data. We thus follow the hierarchical statistical procedure introduced in~\cite{2018NatAs...2..387M} to consecutively add to the ROI model the templates with the highest $\Delta \mathrm{TS}$. As can be seen in Table~\ref{tab:loglike-values}, using this procedure we find that the data strongly supports the inclusion of the NB template first, and subsequently, the BB template. Importantly, in consistency with previous analyses~\citep[e.g.,][]{2018NatAs...2..387M,2018NatAs...2..819B, Macias:2019omb,Abazajian:2020tww}, we find that once the NB and BB templates have been added to the ROI model, the data no longer require any of the DM templates considered in this study. 

\begin{figure}
    \centering
    \includegraphics[width=0.99\linewidth]{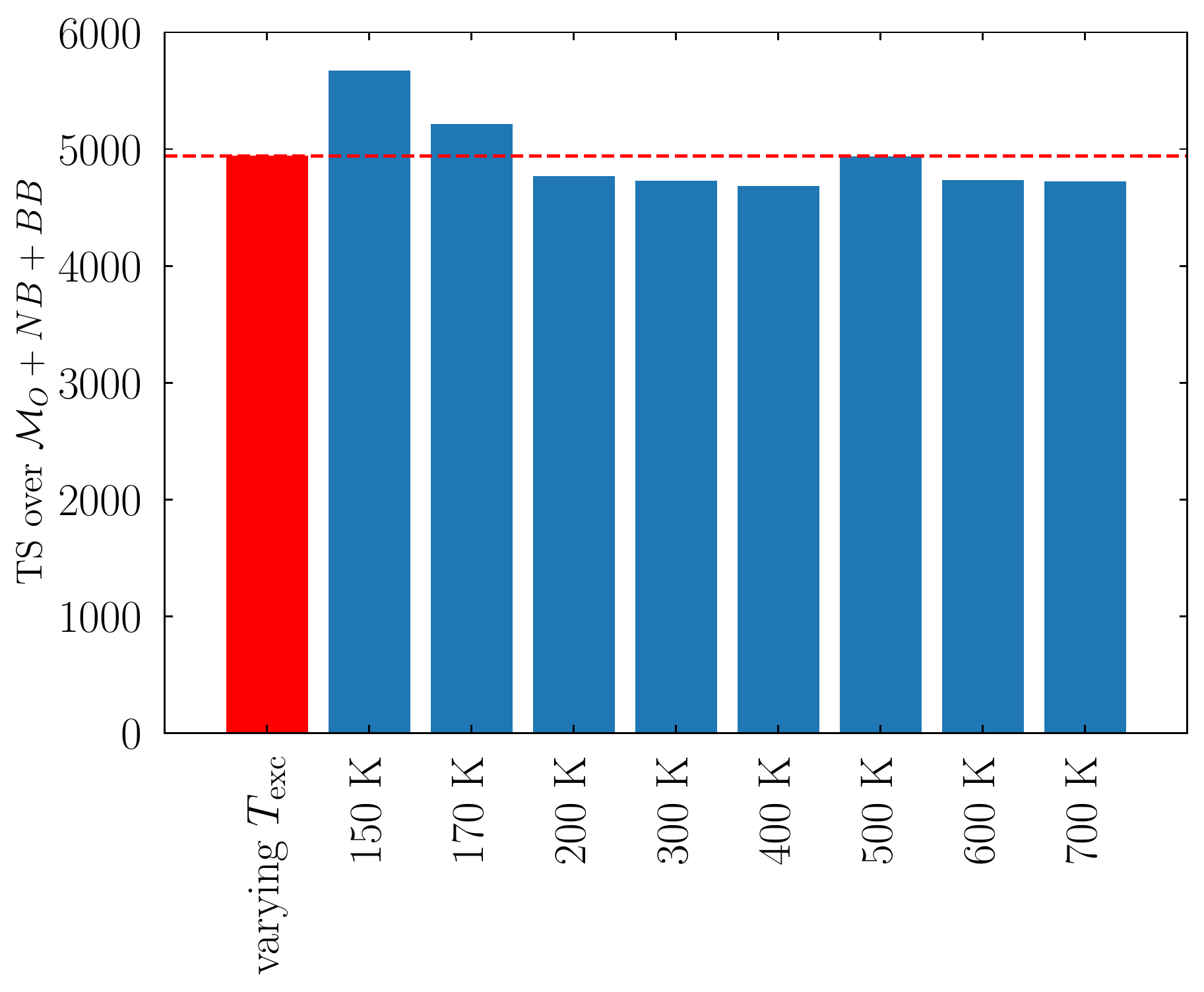}
    \caption{The Test Statistic (TS) of the ``Base+NB+BB'' model in comparison with ``Model O + NB+BB'' that is based on the previous generation of our hydrodynamic gas maps. The blue bars display the results for H$I$ maps with constant $T_{\rm exc}$, and the red bar displays the results for a $T_{\rm exc}$ which varies with $l$ and $b$. Evidently the new H$I$ maps are statistically highly preferred. }
    \label{fig:TSvsExcitationTemp}
\end{figure}

Compared to our previous articles on this subject, we now find a much greater discriminant power for the GCE templates. Our new study shows that the DM templates are statistically highly disfavoured once the NB template is included in the ROI model (see rows 7-11 in Table~\ref{tab:loglike-values}). To be noted from this table is that while the DM templates are strongly disfavoured, the BB template is detected at very high significance ($14.7\sigma$). This highlights that our new hydrodynamic gas maps (with varying $T_{\rm exc}$) drastically improve the sensitivity to the spatial morphology of the GCE templates.


\subsection{{Comparison to hydrodynamic models without continuum emission}}
\label{subsec:comparisonOldGDE}

\cite{2018NatAs...2..387M} demonstrated that the hydrodynamic gas maps introduced in~\cite{2008ApJ...677..283P} give a better fit to the Galactic Center data than other gas models in the literature~\citep[e.g.,][]{2017ApJ...840...43A}. The result has been confirmed with a different methodology in~\cite{Buschmann:2020adf} (see e.g., Fig.~3 in that paper). 

We now investigate whether our new interstellar gas models provide a better fit to the gamma-ray data than those proposed in~\cite{2018NatAs...2..387M}. For this comparison, we use the test statistic, TS, defined as $\mathrm{TS} = 2[\ln(\mathcal{L}_{1})-\ln(\mathcal{L}_{0})]$, where $\mathcal{L}_{1}$  and $\mathcal{L}_{0}$ are the likelihood for the ``Base+NB+BB'' model (see Table~\ref{tab:loglike-values}),  and the ``Model O + NB+BB'' model, respectively. Note that ``Model O'' ($\mathcal{M}_{O}$) assumes the hydrodynamical H$I$ maps introduced in~\cite{2018NatAs...2..387M}, but is otherwise the same as that in the present study.  

Figure~\ref{fig:TSvsExcitationTemp} displays the results of this analysis. Remarkably, all new H$I$ maps are found to have TS values in excess of $ 4500$, and we conclude that our new gas models fit the data significantly better than the previous generation of hydrodynamic gas models \citep{2018NatAs...2..387M,Macias:2019omb,Buschmann:2020adf}. Even though the H$I$ maps with $T_{\rm exc}=150$~K and $170$~K seem to be preferred by the gamma-ray data, they do a poorer job at explaining the radio data (see Fig.~\ref{fig:noise}). Our approach in this study was to select the H$I$ maps which fit the radio data best. These correspond to the H$I$ maps with  $T_{\rm exc}$ varying with $l$ and $b$. We leave for a future investigation to perform a global fit in which we simultaneously fit the radio and gamma-ray data in order to find out the best-fitting excitation temperature $T_{\rm exc}$ for each line of sight.


\subsection{The effect of breaking the Galactic diffuse emission templates into Galactocentric rings.}

\label{comparison with Di Mauro}

Early analyses~\citep[e.g.,][]{Daylan:2014rsa} using astrophysical background models based on GALPROP found that the spatial morphology of the GCE was better modeled by spherically symmetric templates than by templates elongated along (or perpendicular to) the Galactic disk. However, more recent studies using GALPROP maps as well as improved Galactic diffuse emission models~\citep[e.g.,][]{2018NatAs...2..387M,Macias:2019omb, Coleman:2019kax, Abazajian:2020tww} or new statistical methods~\citep[e.g.,][]{2018NatAs...2..819B,Calore:2021jvg} obtained that the GCE data prefers stellar bulge templates rather than a spherically symmetric one. Possibly, the earlier analyses of the GCE did not find evidence for a stellar bulge component, because they did not include an explicit comparison with the stellar bulge templates. The stellar bulge model  has a radially varying asymmetry that is not captured by the elliptical shape tests conducted in, e.g., \citet{Daylan:2014rsa}.  

Recently, the evidence in favor of the stellar bulge explanation of the GCE has been questioned by~\cite{DiMauro:2021raz}, who claimed that dark-matter templates were preferred for all the diffuse emission models considered. However this claim only applies to the combination ``DM+NB'', whereas the ``BB+NB'' combination was better than the ``DM'' template for five out of the seven Galactic diffuse emission models that where considered (see Table 3 of~\cite{DiMauro:2021raz}).
Hence there is little, if any, conflict with, e.g., \citet{2018NatAs...2..387M,2018NatAs...2..819B}, who demonstrated that ``BB+NB'' is better than the ``DM'' (only) hypothesis.
It must be clarified that the stellar bulge model is divided into separate structures (NB+BB) as the NB has a different star formation history compared to the BB~\citep{BlandHawthornandGerhard}, but for consistency both of these templates must be included when considering the Galactic bulge hypothesis. The physical meaning of the ``DM+NB'' hypothesis considered in~\citet{DiMauro:2021raz} is thus unclear.

Nevertheless, \citet{DiMauro:2021raz} kindly provided his Base 
model and thus allowed us to perform an explicit comparison of the morphological results with our method and his. We re-iterate some important differences between the two methods: (i) whereas we divide the gas-correlated gamma-ray maps and IC maps in different concentric rings---which makes our maps less prone to biases---\cite{DiMauro:2021raz} does not use the ring subdivision scheme for the Galactic diffuse emission models, (ii) while our fitting procedure is based on a bin-by-bin method---which allows our results to be independent of the assumed spectra of the background sources---\cite{DiMauro:2021raz} used a broad-band fitting procedure when calculating the TS-values of the sources of interest.  We note that dividing the Galactic diffuse emission models in concentric rings and using a bin-by-bin fitting procedure are the standard methods used by the Fermi-LAT collaboration~\citep{Fermi-LAT:2015sau,2017ApJ...840...43A} for analyses of the GC region. As we show below, the less flexible fitting method assumed in~\cite{DiMauro:2021raz} could explain his different findings.

\begin{figure}
    \centering
    \includegraphics[width=0.99\columnwidth]{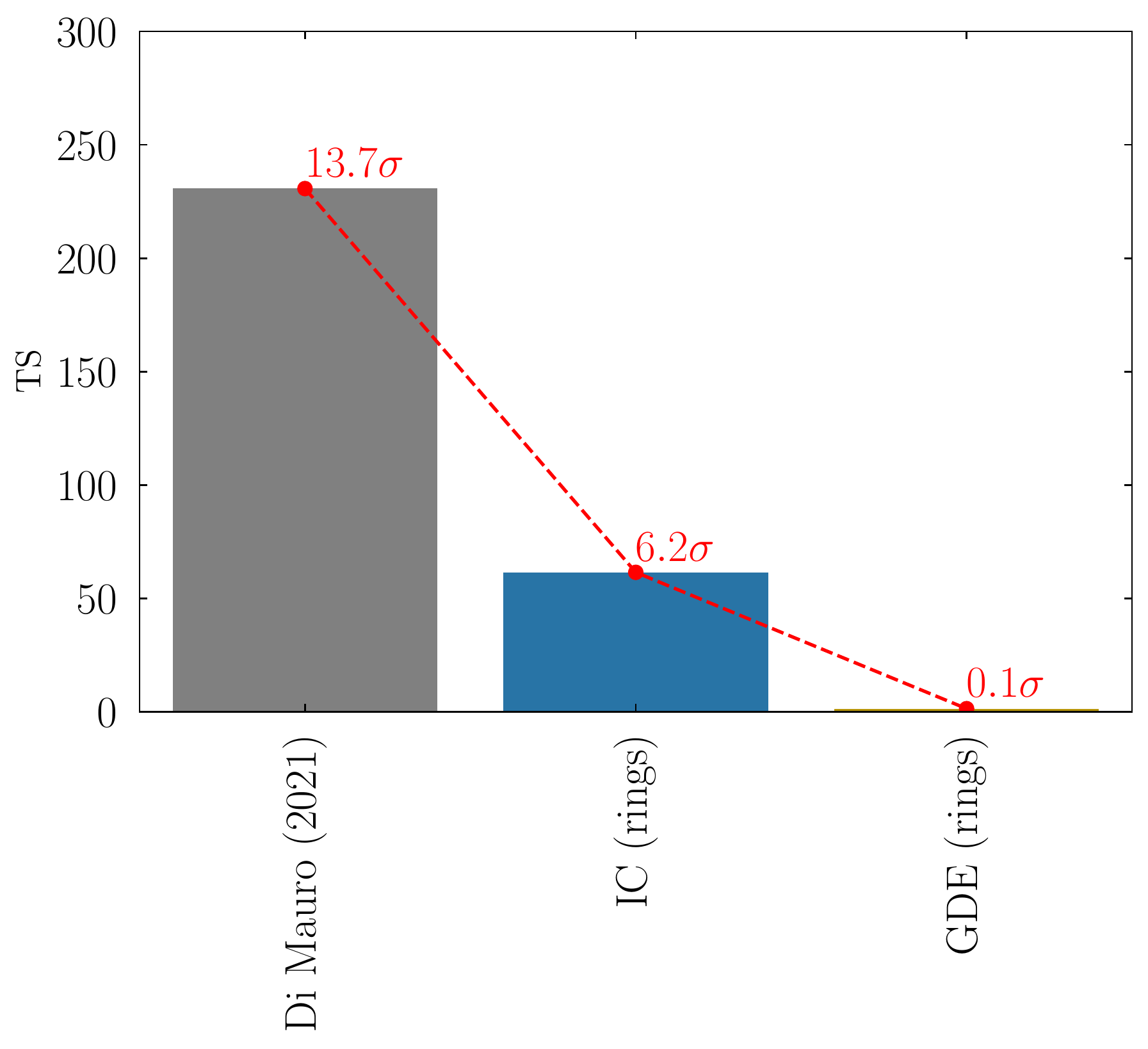}
    \caption{Statistical significance of the DM template over a baseline model that includes the BB+NB templates and the Galactic diffuse emission (GDE) model in \citet{DiMauro:2021raz}. The bars show the TS for the DM template, and the red points the significance in units of sigma for 15 degrees of freedom. The grey bar applies to the Base model of \citep{DiMauro:2021raz}. The blue bar results when we construct divide the IC model of \citep{DiMauro:2021raz} into six concentric rings. The gold bar and $0.1\sigma$ significance for the DM template are found when both the hadronic (four concentric rings) and IC (six concentric rings) components in  \citep{DiMauro:2021raz} are divided into concentric rings. }
    \label{fig:ComparisonDiMauro}
\end{figure}

Comparing the log-likelihood values, we find that our Base model is strongly preferred with $\Delta \mathrm{TS}=21378.4$ over the corresponding Base model in \citet{DiMauro:2021raz}. Already the diffuse emission models based on the earlier generation of hydrodynamic maps \citep{2018NatAs...2..387M,Buschmann:2020adf} are significantly better than those based on the interpolated gas maps in GALPROP. As shown in this paper, our updated hydrodynamic gas maps are yet again a much better fit to the data than are the earlier hydrodynamic gas maps used in \citet{2018NatAs...2..387M}, which is consistent with the very strong preference over the Base model in \citet{DiMauro:2021raz}. 

Secondly, using a bin-by-bin-analysis, the ``Di Mauro Base''+``BB+NB'' provides a better fit to the data with  $\Delta \mathrm{TS}=149.3$ than does ``Di Mauro Base''+``DM'', where the DM template corresponds to a NFW (squared) with $\gamma=1.2$. This shows that even using a Galactic diffuse emission model that gives a poorer fit to the data, the stellar bulge model is preferred. 

Lastly, we constructed IC and hadronic/bremsstrahlung templates that follow closely the GALPROP setup provided in~\cite{DiMauro:2021raz}, but are divided into concentric rings so that the maps have more flexibility to account for potential excesses in the data. We then computed, using a bin-by-bin analysis, the significance of the DM template.
The grey bar in Fig.~\ref{fig:ComparisonDiMauro} shows the statistical significance of the DM template at $13.7\sigma$ for the case in which we have as our baseline model the ``Di Mauro Base'' and BB and NB templates. If we replace the monolithic IC component in Di Mauro's model by an IC divided into six concentric rings, then we observe that the significance of the DM component drops to $6.2\sigma$. This is consistent with earlier work by the Fermi-LAT collaboration~\citep{Fermi-LAT:2015sau} that indicated a large increase in the IC emission in the inner Galaxy, modelling of which requires an IC map divided into concentric rings. Fig.~\ref{fig:ComparisonDiMauro} also shows that the DM significance drops to $0.1\sigma$ when both the IC and the hadronic/bremsstrahlung components are divided into concentric rings.

In summary, we have demonstrated that fitting the highly complex region of the inner Galaxy with inflexible emission models has the potential to create a spurious excess in the data that could resemble a DM signal. Once Galactic diffuse emission models that are divided into concentric rings are included in the fits, the evidence for such an excess disappears.

\begin{figure}
    \centering
    \includegraphics[width=0.99\linewidth]{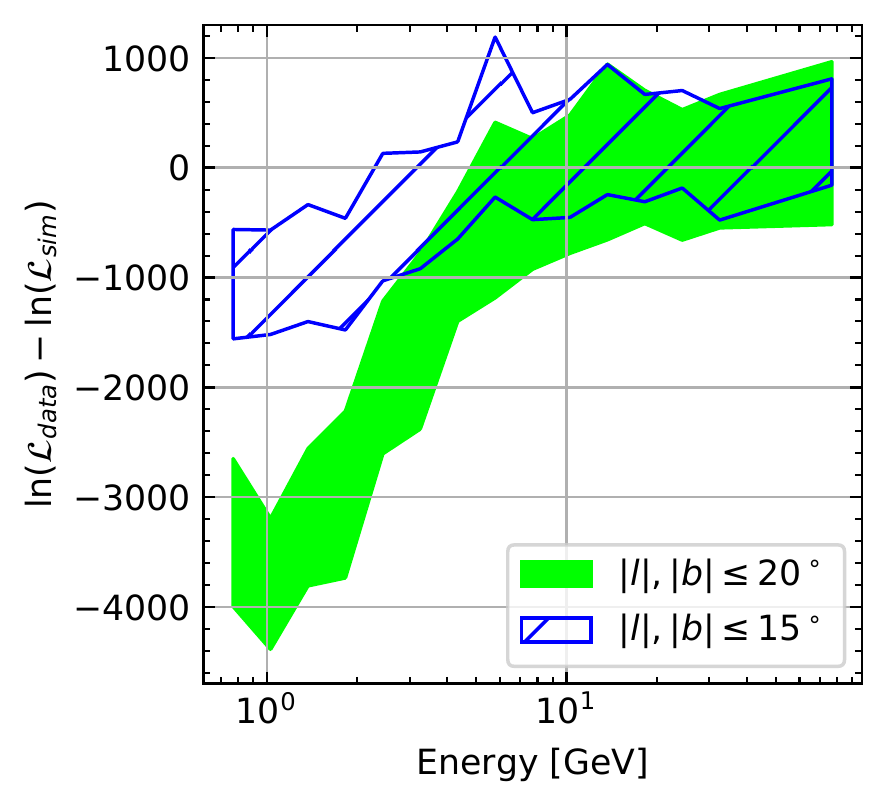}
    \caption{\update{\textbf{Difference in $\ln {\cal L}$ between our best-fitting model (Base+NB+BB),} including the H$I$ maps with $T_{\rm exc}$ varying with $l$ and $b$, applied to the real \textit{Fermi}-LAT data and the same applied to Poissonian Monte Carlo simulations of the best-fit `Base+NB+BB' model.  The green band shows the range of values for 100 MC expectations. The blue hachure indicates the same for an analysis restricted to $\vert l\vert ,\vert b\vert \le 15^\circ$.} 
    }
    \label{fig:goodnessoffit}
\end{figure}

\subsection{Fit validations}

In Sec.~\ref{subsec:implicationsGCE}, we established that the ``Base+NB+BB'' model provides the best fit for our ROI. We now assess whether or not this model is appropriate in an absolute sense, using a similar method to that introduced in~\cite{Buschmann:2020adf}. First, we Monte-Carlo (MC) simulate 100 data sets, each with 15 bands,
drawn from the ``Base+NB+BB'' model---assuming the data are Poisson distributed. Second, we subject these synthetic data sets to exactly the same bin-by-bin fitting procedure as was used on real data. 
{Note that fitting one model to different data is not the same as fitting different models to the same data. Hence we can only find consistency or inconsistency of model and data, indicated by the fit likelihood to the data being, or not being, in the range of likelihoods for the MC data that are drawn from the same model as is used to fit them. Some level of disagreement is to be expected, partly on account of model imperfections, but in particular because we did not run a full scan of sources in the 4FGL and a search for new ones. It is well known that variations in the diffuse-emission model will change the count and the properties of sources, including their location \citep{2015A&A...581A.126S}. 
}

We present the results of the fit validation in Fig.~\ref{fig:goodnessoffit}. {As each MC simulation may have a different number of gamma-ray events in each pixel, $n$, we needed to include the usually dropped $n!$ term in the Poisson likelihood. The marked areas represent the full performance range of the 100 MC models compared to that of the LAT data. To be noted is the large range of $\ln(\cal L)$ values for the simulated data. It exemplifies that when comparing different data to the same model the likelihood is not distributed as for different models adapted to the same data. If the marked area includes the zero level, then the model-based MC data and the LAT data would be statistically consistent.}

{Crudely speaking we expect the true model to be outside of the areas in about 1\% of cases for 100 MC data sets. Fifteen energy bands imply fifteen trials, and so there should be a probability of roughly 15\% to find the true model outside the area in one energy band. Some additional deviation will arise from the limited treatment of point sources, because for expedience we optimized the normalization of only a quarter of the 4FGL sources and varied the location of none of them. Even a perfectly good gas model may therefore give more than one outlier in this test. To remedy this would require refitting the whole Fermi point source catalogue with the new templates and so we postpone it for a future project. Considering the full region $\vert b\vert \le 20^\circ$, our best-fit model is outside of the range of MC results in five energy bands. For $\vert l\vert , \vert b\vert \le 15^\circ$ we see only four outliers below about $2$~GeV, and these stray about a third as far from the zero level with 56\% of the pixel count, indicating a $2.5$ times better fit per pixel at $\vert l\vert , \vert b\vert \le 15^\circ$ than in the outer parts of the ROI. The number of gamma-ray events is highest around a GeV, and so is the sensitivity of this test. We conclude that our model ``Base+NB+BB'' fits reasonably well at $\vert l\vert ,\vert b\vert \le 15^\circ$, where a potential dark-matter signature would matter. There may be some issues at $15^\circ\le (\vert l\vert ,\vert b\vert) \le 20^\circ$ that may have to do with the Fermi bubbles or point sources outside of the ROI affecting the prediction but not being optimized over in the fit.
This} important test gives further credence to our method of separating DM models from those of astrophysical origin based on their spatial morphologies. 

\begin{figure}
    \centering
    \includegraphics[width=0.99\linewidth]{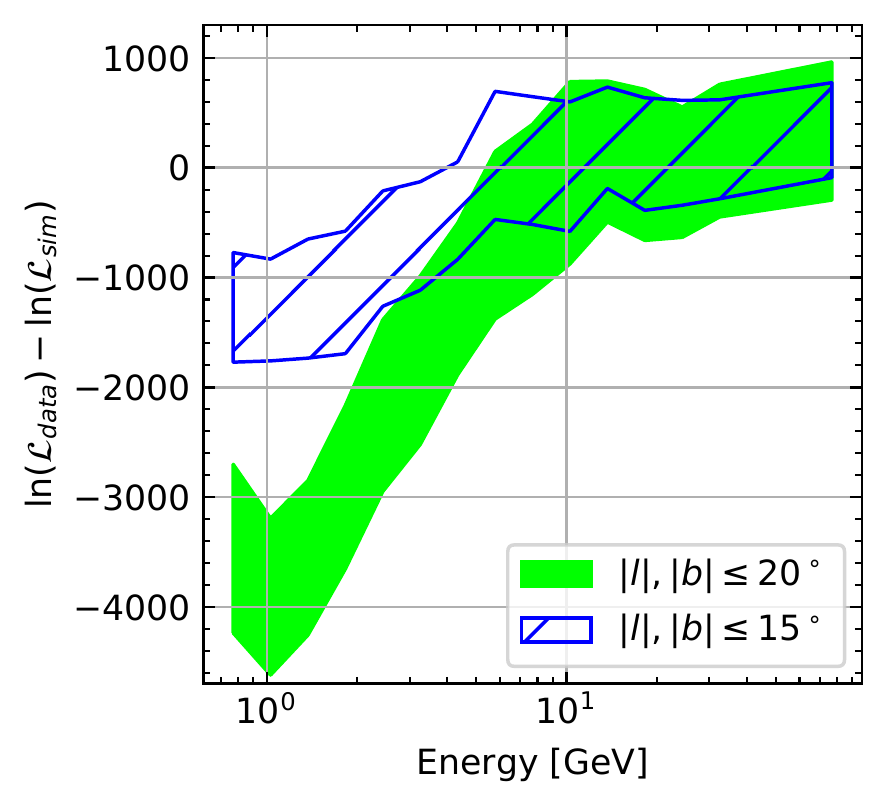}
    \caption{\textbf{Results a fit validation test applied to an Alternative Base model only} that includes H$I$ maps with $T_{\rm exc}=150$ K. The validation tests were done using the same method as in Fig.~\ref{fig:goodnessoffit}. Note that in this case, the NB and BB templates were not included in the fits or the Monte Carlo simulations. }
    \label{fig:alternativegoodnessoffit}
\end{figure}

For comparison purposes, we also performed this same kind of validation test to an ``Alternative Base'' model. This model is defined by substituting the H$I$ maps derived for spatially-varying $T_{\rm exc}$ with those that assume $T_{\rm exc}=150$~K throughout the Galaxy. Notice that for this test, we did not include the NB and BB templates in either the model fits or the simulations.  The results of this exercise are displayed in Fig.~\ref{fig:alternativegoodnessoffit}. {We find the same trend as in Figure~\ref{fig:goodnessoffit}, but a slightly larger discrepancy between the data and the MC expectations in both angular regions. Although the $150$-K gas model provides a good fit to the gamma-ray data (see Figure~\ref{fig:TSvsExcitationTemp}), it is a poor fit to the 21-cm data, as shown by Figure~\ref{fig:noise}. Even if it fits the gamma-ray data almost as well as the model with varying $T_\mathrm{exc}$ and a NB and BB, it is still not a good model}. 

\subsection{Gamma-ray Residuals}

\begin{figure*}[t]
\centering
\includegraphics[width=0.3\textwidth]{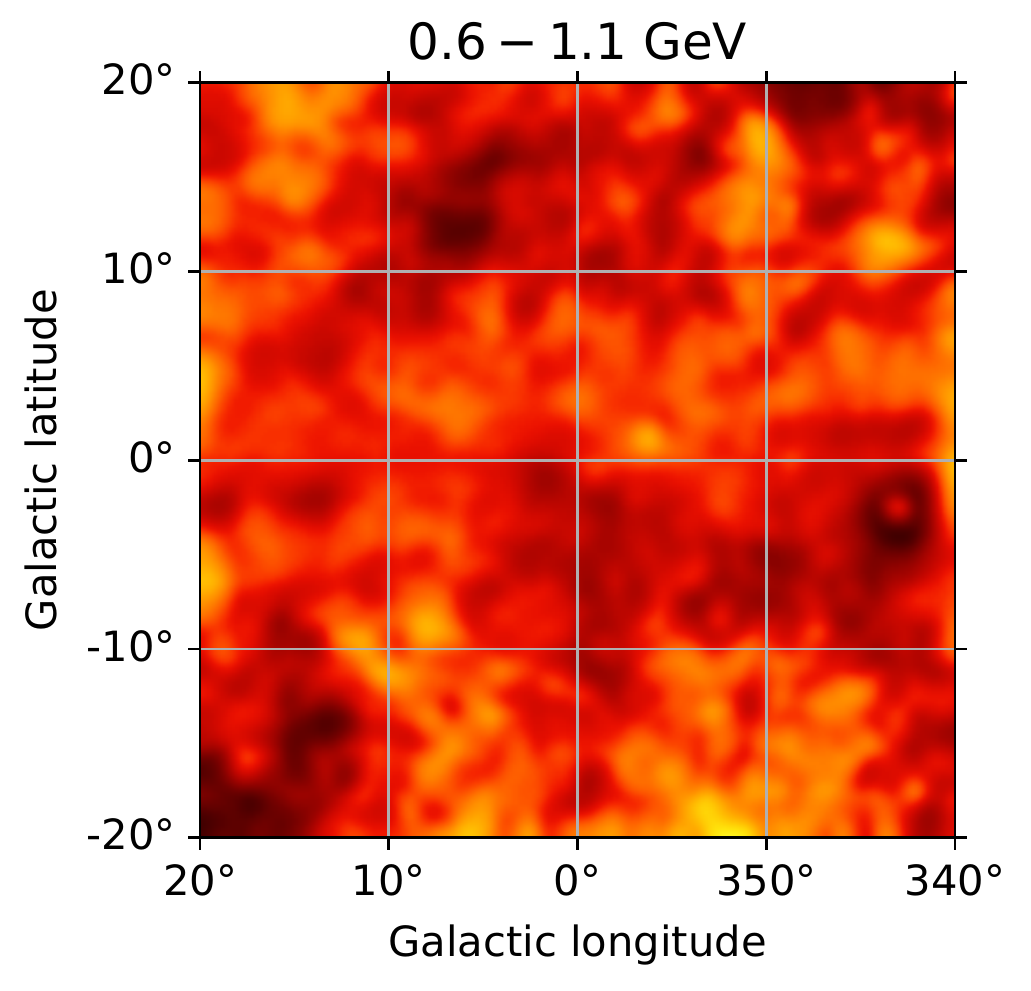} 
 \includegraphics[width=0.3\textwidth]{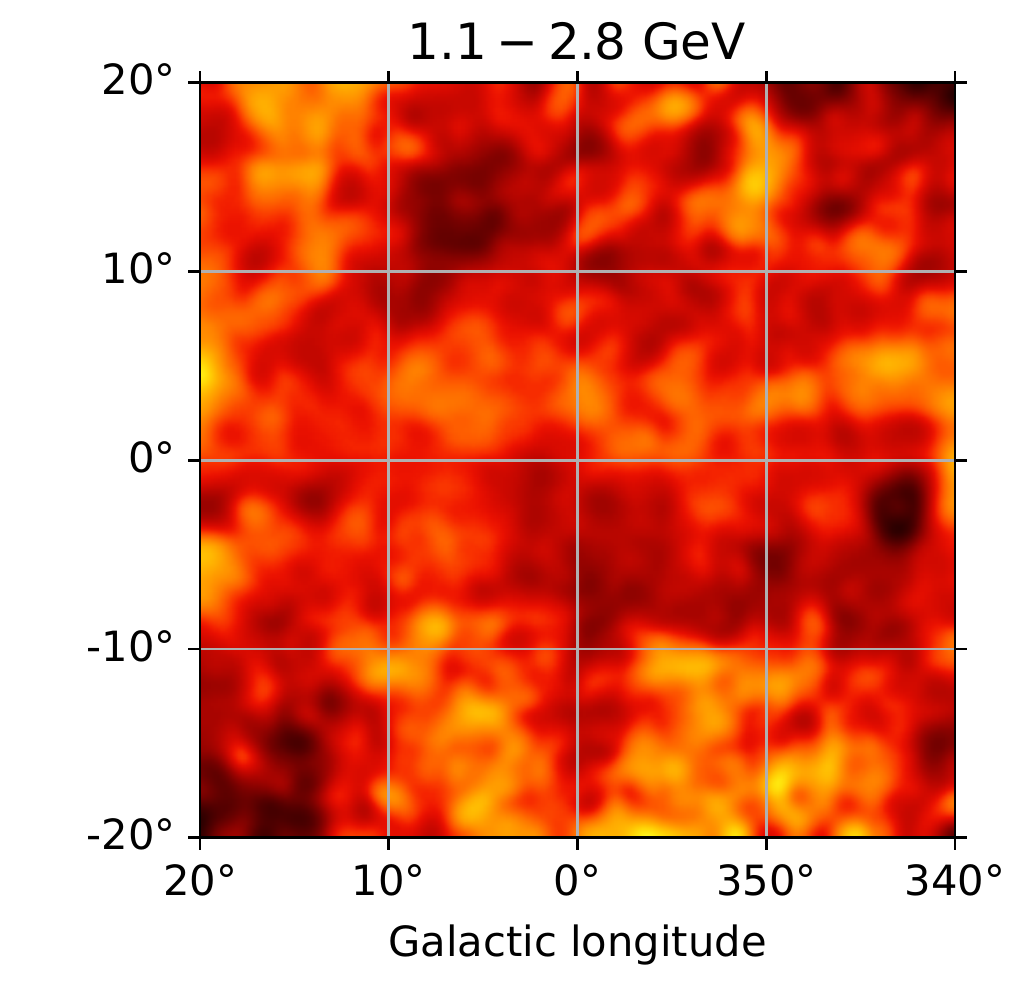} 
 \includegraphics[width=0.37\textwidth]{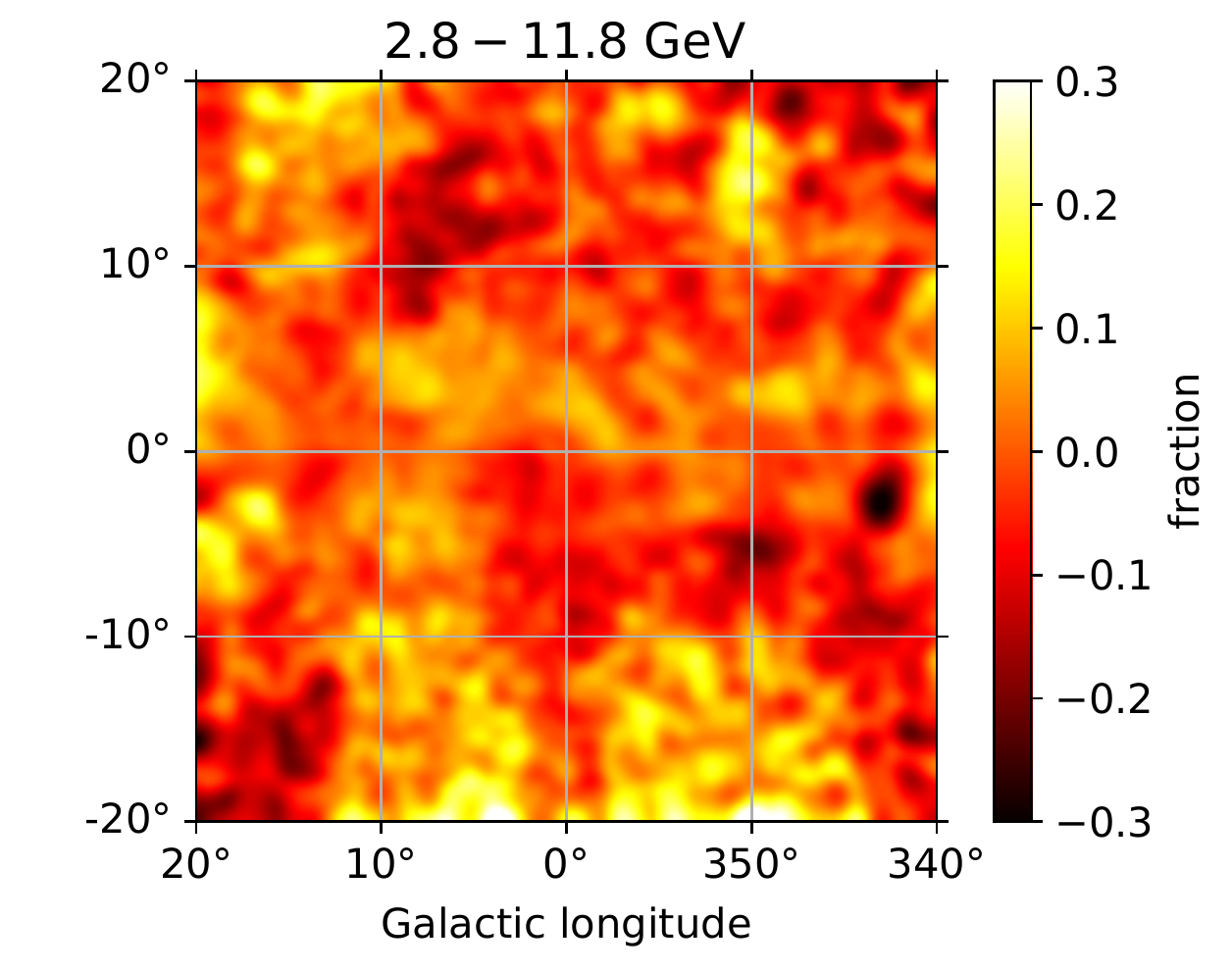}
 \caption{The fractional residuals, (Data-Model)/Model, for the ``Base+NB+BB'' model. Detailed descriptions of the templates included in the ROI model can be found in Appendix~\ref{sec:astrotemplates}, likewise the gamma-ray spectrum for the ``Base+NB+BB'' model.  The images have been smoothed with a Gaussian filter of radius $0.6^\circ$.}
 \label{fig:Residuals}
 \end{figure*}

It is interesting to inspect the residual images for our best-fitting ``Base+NB+BB'' model. Figure~~\ref{fig:Residuals} shows the fractional residuals, $(\rm{Data}-\rm{Model})/\rm{Model}$, in three different macro energy bins: $[0.6,1.1]$, $[1.1,2.8]$ and $[2.8,11.8]$ GeV. They are constructed by joining various micro energy bins, in which the actual fits are performed and that are narrow enough for a reasonable accuracy of our assumption of a flat gamma-ray spectrum in each bin. The mismatch between model and data is mostly at the $\lesssim 10\%$ level, although we also observe localized correlated residuals that reach up to $20-30\%$, {for example one associated with the SNR RX~J1713.7-3946, for which we used the disk template 
provided by the Fermi team in the 4FGL catalog.  }  
These localized residuals are found mostly for latitudes $|b|\gtrsim 8^\circ$   
where the GCE is less significant {and imperfections in the current FB template would leave their signature.} 

{The fit validation tests described in the previous section also indicated a slight deterioration of the fit quality at $\vert l\vert,\vert b\vert \ge 15^\circ$. In the core region $\vert l\vert,\vert b\vert \le 15^\circ$ the validation tests were passed and the residuals are weak, although Figure~\ref{fig:Residuals} shows that our ROI model would still benefit from} further improvements. 
{Possible causes for these localized residuals may include compact sources of radio continuum emission that cannot be accounted for in our radiation-transport model, sub-threshold point sources in the 4FGL catalog which become more statistically significant with our new H$I$ maps, or a suboptimal localization of some 4FGL sources. Another possible cause is that the correlated residuals are due to imperfections in the current FB template. }
Even though our current astrophysical model for the Galactic Center region is not perfect, the fit validation tests do show that the statistical results obtained in our analysis are very robust.
 
\section{Clues about the MSPs formation mechanisms from the morphology of the GCE}

\begin{figure}[t!]
    \centering
    \includegraphics[width=0.99\linewidth]{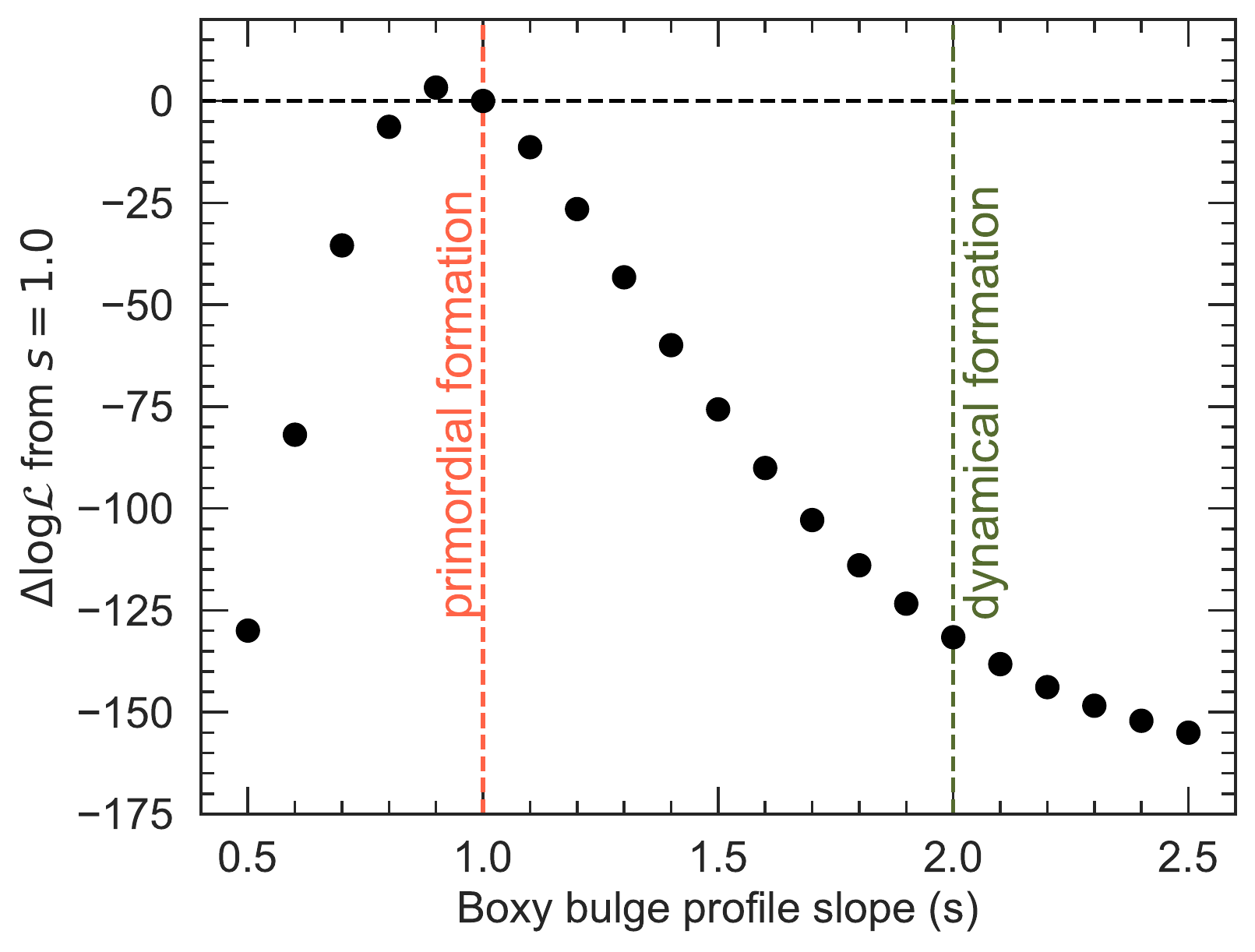}
    \caption{{$\Delta \log(\mathcal{L})$ as a function of the stellar density slope, $s$, with respect to the primordial formation model, for which $s=1$. The GCE data show strong support for scenarios in which the MSPs are formed in situ.} }
    \label{fig:primordial}
\end{figure}

Assuming that unresolved MSPs cause the stellar-bulge feature in the data, \citet{Macias:2019omb} performed morphological tests to determine the evidence for and against the two leading MSP formation scenarios in the literature: (i) the primordial formation, and (ii) dynamical formation. Galactic bulge templates were created based on the boxy bulge stellar density, $\rho^s_{\rm bar} (R,\phi,z)$, where $R$, $\phi$, $z$ are the cylindrical coordinates, and $s$ is a profile slope. In the primordial (dynamical) formation scenario it is expected that the GCE data matches the stellar density template with slope $s=1$ ($s=2$). Using the previous generation~\citep{2018NatAs...2..387M} of hydrodynamic gas models, \cite{Macias:2019omb} found a best-fit slope $s=1.38^{+0.06}_{-0.05}$ with a $\Delta \log(\mathcal{L})=28.8$ for one new parameter, suggesting that some MSPs are formed in situ and others through stellar interactions. 

Figure~\ref{fig:primordial} shows the updated $\Delta \log(\mathcal{L})$ as a function of the stellar-density slope, $s$, with respect to the primordial formation model [cf. Fig.9c in \citep{Macias:2019omb}]. To be noted from the figure is a strong preference for $s\approx 1$ with the updated Galactic diffuse emission model. In other words, these new results point to a formation scenario that is consistent with the Galactic bulge MSPs being formed in situ.

\section{Conclusions}

We devised a new model of the Galactic distribution of atomic hydrogen, H$I$, that traces gas both in emission and in absorption. For that purpose, we constructed a model of continuum emissivity in the 21-cm waveband that reproduces the continuum brightness observed from the inner Galaxy with the CHIPASS survey \citep{2014PASA...31....7C}. We then solved the radiation transport integral for H$I$ line emission in the presence of continuum radiation for each line of sight and Doppler shift, for which 
the H$I$4PI survey reports a non-zero line signal. The mapping of Doppler shift to distance along the line of sight is performed using the gas-flow model of \citet{2003MNRAS.340..949B} and the algorithm described in \citet{2008ApJ...677..283P} that provide distance resolution also toward the Galactic Center. 

Our explicit radiation-transport modelling can reproduce the negative line signals that one often finds within a few degrees off the Galactic Center. Atomic gas seen in absorption can thus be accounted for. We find an enhanced column density attributed to the Galactic plane at $r\le 3.5$~kpc, where H$I$ absorption is particularly strong. Within a few degrees off the Galactic Center, this signal is not simply taken from other locations on the line of sight. Instead it results from the proper modelling of H$I$ absorption and the strong continuum emission from that direction. 

We test various values of the hydrogen excitation temperature, $T_\mathrm{exc}$, ranging from $130$~K to $700$~K. The lower the excitation temperature, the easier it is to reproduce absorption features in the spectra, in particular negative line signals. At the same time, one cannot model line signal with brightness temperatures exceeding $T_\mathrm{exc}$, and in the presence of continuum emission the achievable line brightness can be well below this limit. For a constant excitation temperature, we find that the H$I$4PI spectra are best reproduced for $T_\mathrm{exc}=200$~K with an average mismatch below $0.08$~K or about twice the survey sensitivity. The mismatch increases slowly for higher excitation temperatures and does so quite rapidly for $T_\mathrm{exc}\lesssim 170$~K. We also constructed a model of the Galactic distribution of atomic hydrogen, in which we allowed $T_\mathrm{exc}$ to vary as a function of $l$ and $b$. This model fits the line data best and serves as a fiducial model for the subsequent analysis of the diffuse gamma-ray emission from the inner Galaxy.

We then updated our model of the diffuse gamma-ray emission from the inner Galaxy \citep{2018NatAs...2..387M} with the new maps of Galactic atomic hydrogen and  new templates for the dust correction. The model comprises components that describe cosmic-ray induced gamma-ray emission, large-scale features like the Fermi bubbles, a nuclear bulge and a boxy bulge, and minor aspects like the Sun and the Moon. The new H$I$ map affects the cosmic-ray induced gamma-ray emission through hadronic interactions and nonthermal bremsstrahlung. We find with high significance, $\Delta \mathrm{TS}\approx 5000$, a much better fit to the diffuse gamma-ray emission from the inner $40^\circ \times 40^\circ $ of the Galaxy as observed with the \textit{Fermi}-LAT, if our new H$I$ model is used. A similar improvement in fit quality is seen for all choices of $T_\mathrm{exc}$ that we probed. The likelihood fit still requires that templates for the nuclear bulge \citep{Nishiyama2015} and the boxy bulge \citep{Coleman:2019kax} are included in the model, as was the case in earlier analyses. Already without the boxy bulge, but also with it, there is no evidence for any of the dark-matter scenarios we tested. These include with arbitrary spectral form cuspy and cored dark-matter profiles and ellipsoidal versions thereof.

We performed various checks for potential systematic issues without finding an indication for any. The results appear to be robust. Compared to previous studies, we now find a much greater discriminant power for the templates for the Galactic-Center excess. While the dark-matter templates do not significantly improve the fit, the boxy bulge template is detected at  nearly the 15$\sigma$ level. We conclude that our new hydrodynamic gas maps, allowing $T_\mathrm{exc}$ to vary as a function of $l$ and $b$, not only provide an unprecedented reconstruction of H$I$ line spectra, but also drastically improve the sensitivity to the spatial morphology of the various components of diffuse Galactic gamma-ray emission for the much-discussed  Galactic Center excess.

We foresee that our new H$I$ maps will be very useful for the ambitious Galactic Center survey program of the forthcoming Cherenkov Telescope Array~\citep{CTA:2020qlo} and particularly for characterizing the high-energy tail of the GCE at TeV-scale energies~\citep{Song:2019nrx,Macias:2021boz}.

\section{Acknowledgments} 
We thank Shin'ichiro Ando, Roland M. Crocker, and Shunsaku Horiuchi for fruitful discussions. We are also grateful with Mattia Di Mauro for sharing his Base Galactic diffuse emission model with us for comparisons with his results. O.M. is supported by the GRAPPA Prize Fellowship.

\bibliography{refs}{}

\begin{thebibliography}{}
\expandafter\ifx\csname natexlab\endcsname\relax\def\natexlab#1{#1}\fi
\providecommand{\url}[1]{\href{#1}{#1}}
\providecommand{\dodoi}[1]{doi:~\href{http://doi.org/#1}{\nolinkurl{#1}}}
\providecommand{\doeprint}[1]{\href{http://ascl.net/#1}{\nolinkurl{http://ascl.net/#1}}}
\providecommand{\doarXiv}[1]{\href{https://arxiv.org/abs/#1}{\nolinkurl{https://arxiv.org/abs/#1}}}

\bibitem[{{Abazajian} {et~al.}(2020){Abazajian}, {Horiuchi}, {Kaplinghat},
  {Keeley}, \& {Macias}}]{Abazajian:2020tww}
{Abazajian}, K.~N., {Horiuchi}, S., {Kaplinghat}, M., {Keeley}, R.~E., \&
  {Macias}, O. 2020, \prd, 102, 043012, \dodoi{10.1103/PhysRevD.102.043012}

\bibitem[{{Abdollahi} {et~al.}(2020){Abdollahi}, {Acero}, {Ackermann},
  {Ajello}, {Atwood}, {Axelsson}, {Baldini}, {Ballet}, {Barbiellini},
  {Bastieri}, {Becerra Gonzalez}, {Bellazzini}, {Berretta}, {Bissaldi},
  {Blandford}, {Bloom}, {Bonino}, {Bottacini}, {Brandt}, {Bregeon}, {Bruel},
  {Buehler}, {Burnett}, {Buson}, {Cameron}, {Caputo}, {Caraveo}, {Casandjian},
  {Castro}, {Cavazzuti}, {Charles}, {Chaty}, {Chen}, {Cheung}, {Chiaro},
  {Ciprini}, {Cohen-Tanugi}, {Cominsky}, {Coronado-Bl{\'a}zquez}, {Costantin},
  {Cuoco}, {Cutini}, {D'Ammando}, {DeKlotz}, {de la Torre Luque}, {de Palma},
  {Desai}, {Digel}, {Di Lalla}, {Di Mauro}, {Di Venere}, {Dom{\'\i}nguez},
  {Dumora}, {Fana Dirirsa}, {Fegan}, {Ferrara}, {Franckowiak}, {Fukazawa},
  {Funk}, {Fusco}, {Gargano}, {Gasparrini}, {Giglietto}, {Giommi}, {Giordano},
  {Giroletti}, {Glanzman}, {Green}, {Grenier}, {Griffin}, {Grondin}, {Grove},
  {Guiriec}, {Harding}, {Hayashi}, {Hays}, {Hewitt}, {Horan},
  {J{\'o}hannesson}, {Johnson}, {Kamae}, {Kerr}, {Kocevski}, {Kovac'evic'},
  {Kuss}, {Landriu}, {Larsson}, {Latronico}, {Lemoine-Goumard}, {Li},
  {Liodakis}, {Longo}, {Loparco}, {Lott}, {Lovellette}, {Lubrano}, {Madejski},
  {Maldera}, {Malyshev}, {Manfreda}, {Marchesini}, {Marcotulli},
  {Mart{\'\i}-Devesa}, {Martin}, {Massaro}, {Mazziotta}, {McEnery}, {Mereu},
  {Meyer}, {Michelson}, {Mirabal}, {Mizuno}, {Monzani}, {Morselli},
  {Moskalenko}, {Negro}, {Nuss}, {Ojha}, {Omodei}, {Orienti}, {Orlando},
  {Ormes}, {Palatiello}, {Paliya}, {Paneque}, {Pei}, {Pe{\~n}a-Herazo},
  {Perkins}, {Persic}, {Pesce-Rollins}, {Petrosian}, {Petrov}, {Piron}, {Poon},
  {Porter}, {Principe}, {Rain{\`o}}, {Rando}, {Razzano}, {Razzaque}, {Reimer},
  {Reimer}, {Remy}, {Reposeur}, {Romani}, {Saz Parkinson}, {Schinzel},
  {Serini}, {Sgr{\`o}}, {Siskind}, {Smith}, {Spandre}, {Spinelli}, {Strong},
  {Suson}, {Tajima}, {Takahashi}, {Tak}, {Thayer}, {Thompson}, {Tibaldo},
  {Torres}, {Torresi}, {Valverde}, {Van Klaveren}, {van Zyl}, {Wood},
  {Yassine}, \& {Zaharijas}}]{Fermi-LAT:4FGL}
{Abdollahi}, S., {Acero}, F., {Ackermann}, M., {et~al.} 2020, \apjs, 247, 33,
  \dodoi{10.3847/1538-4365/ab6bcb}

\bibitem[{{Acharyya} {et~al.}(2021){Acharyya}, {Adam}, {Adams}, {Agudo},
  {Aguirre-Santaella}, {Alfaro}, {Alfaro}, {Alispach}, {Aloisio}, {Alves
  Batista}, {Amati}, {Ambrosi}, {Ang{\"u}ner}, {Antonelli}, {Aramo}, {Araudo},
  {Armstrong}, {Arqueros}, {Asano}, {Ascas{\'\i}bar}, {Ashley}, {Balazs},
  {Ballester}, {Baquero Larriva}, {Barbosa Martins}, {Barkov}, {Barres de
  Almeida}, {Barrio}, {Bastieri}, {Becerra}, {Beck}, {Becker Tjus}, {Benbow},
  {Benito}, {Berge}, {Bernardini}, {Bernl{\"o}hr}, {Berti}, {Bertucci},
  {Beshley}, {Biasuzzi}, {Biland}, {Bissaldi}, {Biteau}, {Blanch}, {Blazek},
  {Bocchino}, {Boisson}, {Bonneau Arbeletche}, {Bordas}, {Bosnjak},
  {Bottacini}, {Bozhilov}, {Bregeon}, {Brill}, {Bringmann}, {Brown}, {Brun},
  {Brun}, {Bruno}, {Bulgarelli}, {Burton}, {Burtovoi}, {Buscemi}, {Cameron},
  {Capasso}, {Caproni}, {Capuzzo-Dolcetta}, {Caraveo}, {Carosi}, {Carosi},
  {Casanova}, {Cascone}, {Cassol}, {Catalani}, {Cauz}, {Cerruti}, {Chadwick},
  {Chaty}, {Chen}, {Chernyakova}, {Chiaro}, {Chiavassa}, {Chikawa}, {Chudoba},
  {{\c{C}}olak}, {Conforti}, {Coniglione}, {Conte}, {Contreras},
  {Coronado-Blazquez}, {Costa}, {Costantini}, {Cotter}, {Cristofari},
  {D'Aimath}, {D'Ammando}, {Damone}, {Daniel}, {Dazzi}, {De Angelis}, {De
  Caprio}, {de C{\'a}ssia dos Anjos}, {de Gouveia Dal Pino}, {De Lotto}, {De
  Martino}, {de O{\~n}a Wilhelmi}, {De Palma}, {de Souza}, {Delgado}, {Delgado
  Giler}, {della Volpe}, {Depaoli}, {Di Girolamo}, {Di Pierro}, {Di Venere},
  {Diebold}, {Dmytriiev}, {Dom{\'\i}nguez}, {Donini}, {Doro}, {Ebr}, {Eckner},
  {Edwards}, {Ekoume}, {Els{\"a}sser}, {Evoli}, {Falceta-Goncalves},
  {Fedorova}, {Fegan}, {Feng}, {Ferrand}, {Ferrara}, {Fiandrini}, {Fiasson},
  {Filipovic}, {Fioretti}, {Fiori}, {Foffano}, {Fontaine}, {Fornieri},
  {Franco}, {Fukami}, {Fukui}, {Gaggero}, {Galaz}, {Gammaldi}, {Garcia},
  {Garczarczyk}, {Gascon}, {Gent}, {Ghalumyan}, {Gianotti}, {Giarrusso},
  {Giavitto}, {Giglietto}, {Giordano}, {Giuliani}, {Glicenstein}, {Gnatyk},
  {Goldoni}, {Gonz{\'a}lez}, {Gourgouliatos}, {Granot}, {Grasso}, {Green},
  {Grillo}, {Gueta}, {Gunji}, {Halim}, {Hassan}, {Heller}, {Hern{\'a}ndez
  Cadena}, {Hiroshima}, {Hnatyk}, {Hofmann}, {Holder}, {Horan}, {H{\"o}randel},
  {Horvath}, {Hovatta}, {Hrabovsky}, {Hrupec}, {Hughes}, {Humensky},
  {H{\"u}tten}, {Iarlori}, {Inada}, {Inoue}, {Iocco}, {Iori}, {Jamrozy},
  {Janecek}, {Jin}, {Jouvin}, {Jurysek}, {Karukes}, {Katarzy{\'n}ski},
  {Kazanas}, {Kerszberg}, {Kherlakian}, {Kissmann}, {Kn{\"o}dlseder},
  {Kobayashi}, {Kohri}, {Komin}, {Kubo}, {Kushida}, {Lamanna}, {Lapington},
  {Laporte}, {Leigui de Oliveira}, {Lenain}, {Leone}, {Leto}, {Lindfors},
  {Lohse}, {Lombardi}, {Longo}, {Lopez}, {L{\'o}pez}, {L{\'o}pez-Coto},
  {Loporchio}, {Luque-Escamilla}, {Mach}, {Maggio}, {Maier}, {Mallamaci},
  {Malta Nunes de Almeida}, {Mandat}, {Manganaro}, {Mangano}, {Manic{\`o}},
  {Marculewicz}, {Mariotti}, {Markoff}, {Marquez}, {Mart{\'\i}}, {Martinez},
  {Mart{\'\i}nez}, {Mart{\'\i}nez}, {Mart{\'\i}nez-Huerta}, {Maurin}, {Mazin},
  {Mbarubucyeye}, {Medina Miranda}, {Meyer}, {Miceli}, {Miener}, {Minev},
  {Miranda}, {Mirzoyan}, {Mizuno}, {Mode}, {Moderski}, {Mohrmann}, {Molina},
  {Montaruli}, {Moralejo}, {Morcuende-Parrilla}, {Morselli}, {Mukherjee},
  {Mundell}, {Nagai}, {Nakamori}, {Nemmen}, {Niemiec}, {Nieto}, {Niko{\l}ajuk},
  {Ninci}, {Noda}, {Nosek}, {Nozaki}, {Ohira}, {Ohishi}, {Ohtani}, {Oka},
  {Okumura}, {Ong}, {Orienti}, {Orito}, {Orlandini}, {Orlando}, {Orlando},
  {Ostrowski}, {Oya}, {Pagano}, {Pagliaro}, {Palatiello}, {Pantaleo},
  {Paredes}, {Pareschi}, {Parmiggiani}, {Patricelli}, {Pavleti{\'c}}, {Pe'er},
  {Pecimotika}, {P{\'e}rez-Romero}, {Persic}, {Petruk}, {Pfrang}, {Piano},
  {Piatteli}, {Pietropaolo}, {Pillera}, {Pilszyk}, {Pintore}, {Pohl},
  {Poireau}, {Prado}, {Prandini}, {Prast}, {Principe}, {Prokoph}, {Prouza},
  {Przybilski}, {P{\"u}hlhofer}, {Pumo}, {Queiroz}, {Quirrenbach}, {Rain{\`o}},
  {Rando}, {Razzaque}, {Recchia}, {Reimer}, {Reisenegger}, {Renier}, {Rhode},
  {Ribeiro}, {Rib{\'o}}, {Richtler}, {Rico}, {Rieger}, {Rinchiuso}, {Rizi},
  {Rodriguez}, {Rodriguez Fernandez}, {Rodriguez Ramirez}, {Rojas}, {Romano},
  {Romeo}, {Rosado}, {Rowell}, {Rudak}, {Russo}, {Sadeh}, {S{\ae}ther Hatlen},
  {Safi-Harb}, {Salesa Greus}, {Salina}, {Sanchez}, {S{\'a}nchez-Conde},
  {Sangiorgi}, {Sano}, {Santander}, {Santos}, {Santos-Lima}, {Sanuy}, {Sarkar},
  {Saturni}, {Sawangwit}, {Schussler}, {Schwanke}, {Sciacca}, {Scuderi},
  {Seglar-Arroyo}, {Sergijenko}, {Servillat}, {Seweryn}, {Shalchi}, {Sharma},
  {Shellard}, {Siejkowski}, {Silk}, {Siqueira}, {Sliusar}, {S{\l}owikowska},
  {Sokolenko}, {Sol}, {Spencer}, {Stamerra}, {Stani{\v{c}}}, {Starling},
  {Stolarczyk}, {Straumann}, {Stri{\v{s}}kovi{\'c}}, {Suda}, {Suomijarvi},
  {{\'S}wierk}, {Tavecchio}, {Taylor}, {Tejedor}, {Teshima}, {Testa},
  {Tibaldo}, {Todero Peixoto}, {Tokanai}, {Tonev}, {Tosti}, {Tosti}, {Tothill},
  {Truzzi}, {Travnicek}, {Vagelli}, {Vallage}, {Vallania}, {van Eldik},
  {Vandenbroucke}, {Varner}, {Vassiliev}, {V{\'a}zquez Acosta}, {Vecchi},
  {Ventura}, {Vercellone}, {Vergani}, {Verna}, {Viana}, {Vigorito}, {Vink},
  {Vitale}, {Vorobiov}, {Vovk}, {Vuillaume}, {Wagner}, {Walter}, {Watson},
  {Weniger}, {White}, {White}, {Wiemann}, {Wierzcholska}, {Will}, {Williams},
  {Wischnewski}, {Yanagita}, {Yang}, {Yoshikoshi}, {Zacharias}, {Zaharijas},
  {Zakaria}, {Zampieri}, {Zanin}, {Zaric}, {Zavrtanik}, {Zavrtanik},
  {Zdziarski}, {Zech}, {Zechlin}, {Zhdanov}, \& {{\v{Z}}ivec}}]{CTA:2020qlo}
{Acharyya}, A., {Adam}, R., {Adams}, C., {et~al.} 2021, \jcap, 2021, 057,
  \dodoi{10.1088/1475-7516/2021/01/057}

\bibitem[{{Ackermann} {et~al.}(2014){Ackermann}, {Albert}, {Atwood}, {Baldini},
  {Ballet}, {Barbiellini}, {Bastieri}, {Bellazzini}, {Bissaldi}, {Blandford},
  {Bloom}, {Bottacini}, {Brandt}, {Bregeon}, {Bruel}, {Buehler}, {Buson},
  {Caliandro}, {Cameron}, {Caragiulo}, {Caraveo}, {Cavazzuti}, {Cecchi},
  {Charles}, {Chekhtman}, {Chiang}, {Chiaro}, {Ciprini}, {Claus},
  {Cohen-Tanugi}, {Conrad}, {Cutini}, {D'Ammando}, {de Angelis}, {de Palma},
  {Dermer}, {Digel}, {Di Venere}, {Silva}, {Drell}, {Favuzzi}, {Ferrara},
  {Focke}, {Franckowiak}, {Fukazawa}, {Funk}, {Fusco}, {Gargano}, {Gasparrini},
  {Germani}, {Giglietto}, {Giordano}, {Giroletti}, {Godfrey}, {Gomez-Vargas},
  {Grenier}, {Guiriec}, {Hadasch}, {Harding}, {Hays}, {Hewitt}, {Hou},
  {Jogler}, {J{\'o}hannesson}, {Johnson}, {Johnson}, {Kamae}, {Kataoka},
  {Kn{\"o}dlseder}, {Kocevski}, {Kuss}, {Larsson}, {Latronico}, {Longo},
  {Loparco}, {Lovellette}, {Lubrano}, {Malyshev}, {Manfreda}, {Massaro},
  {Mayer}, {Mazziotta}, {McEnery}, {Michelson}, {Mitthumsiri}, {Mizuno},
  {Monzani}, {Morselli}, {Moskalenko}, {Murgia}, {Nemmen}, {Nuss}, {Ohsugi},
  {Omodei}, {Orienti}, {Orlando}, {Ormes}, {Paneque}, {Panetta}, {Perkins},
  {Pesce-Rollins}, {Petrosian}, {Piron}, {Pivato}, {Rain{\`o}}, {Rando},
  {Razzano}, {Razzaque}, {Reimer}, {Reimer}, {S{\'a}nchez-Conde}, {Schaal},
  {Schulz}, {Sgr{\`o}}, {Siskind}, {Spandre}, {Spinelli}, {Stawarz}, {Strong},
  {Suson}, {Tahara}, {Takahashi}, {Thayer}, {Tibaldo}, {Tinivella}, {Torres},
  {Tosti}, {Troja}, {Uchiyama}, {Vianello}, {Werner}, {Winer}, {Wood}, {Wood},
  \& {Zaharijas}}]{Fermi-LAT:2014sfa}
{Ackermann}, M., {Albert}, A., {Atwood}, W.~B., {et~al.} 2014, \apj, 793, 64,
  \dodoi{10.1088/0004-637X/793/1/64}

\bibitem[{{Ackermann} {et~al.}(2017){Ackermann}, {Ajello}, {Albert}, {Atwood},
  {Baldini}, {Ballet}, {Barbiellini}, {Bastieri}, {Bellazzini}, {Bissaldi},
  {Blandford}, {Bloom}, {Bonino}, {Bottacini}, {Brandt}, {Bregeon}, {Bruel},
  {Buehler}, {Burnett}, {Cameron}, {Caputo}, {Caragiulo}, {Caraveo},
  {Cavazzuti}, {Cecchi}, {Charles}, {Chekhtman}, {Chiang}, {Chiappo}, {Chiaro},
  {Ciprini}, {Conrad}, {Costanza}, {Cuoco}, {Cutini}, {D'Ammando}, {de Palma},
  {Desiante}, {Digel}, {Di Lalla}, {Di Mauro}, {Di Venere}, {Drell}, {Favuzzi},
  {Fegan}, {Ferrara}, {Focke}, {Franckowiak}, {Fukazawa}, {Funk}, {Fusco},
  {Gargano}, {Gasparrini}, {Giglietto}, {Giordano}, {Giroletti}, {Glanzman},
  {Gomez-Vargas}, {Green}, {Grenier}, {Grove}, {Guillemot}, {Guiriec},
  {Gustafsson}, {Harding}, {Hays}, {Hewitt}, {Horan}, {Jogler}, {Johnson},
  {Kamae}, {Kocevski}, {Kuss}, {La Mura}, {Larsson}, {Latronico}, {Li},
  {Longo}, {Loparco}, {Lovellette}, {Lubrano}, {Magill}, {Maldera}, {Malyshev},
  {Manfreda}, {Martin}, {Mazziotta}, {Michelson}, {Mirabal}, {Mitthumsiri},
  {Mizuno}, {Moiseev}, {Monzani}, {Morselli}, {Negro}, {Nuss}, {Ohsugi},
  {Orienti}, {Orlando}, {Ormes}, {Paneque}, {Perkins}, {Persic},
  {Pesce-Rollins}, {Piron}, {Principe}, {Rain{\`o}}, {Rando}, {Razzano},
  {Razzaque}, {Reimer}, {Reimer}, {S{\'a}nchez-Conde}, {Sgr{\`o}}, {Simone},
  {Siskind}, {Spada}, {Spandre}, {Spinelli}, {Suson}, {Tajima}, {Tanaka},
  {Thayer}, {Tibaldo}, {Torres}, {Troja}, {Uchiyama}, {Vianello}, {Wood},
  {Wood}, {Zaharijas}, {Zimmer}, \& {Fermi LAT
  Collaboration}}]{2017ApJ...840...43A}
{Ackermann}, M., {Ajello}, M., {Albert}, A., {et~al.} 2017, \apj, 840, 43,
  \dodoi{10.3847/1538-4357/aa6cab}

\bibitem[{Ajello {et~al.}(2016)}]{Fermi-LAT:2015sau}
Ajello, M., {et~al.} 2016, Astrophys. J., 819, 44,
  \dodoi{10.3847/0004-637X/819/1/44}

\bibitem[{{Baba} {et~al.}(2010){Baba}, {Saitoh}, \&
  {Wada}}]{2010PASJ...62.1413B}
{Baba}, J., {Saitoh}, T.~R., \& {Wada}, K. 2010, \pasj, 62, 1413,
  \dodoi{10.1093/pasj/62.6.1413}

\bibitem[{Bartels {et~al.}(2016)Bartels, Krishnamurthy, \&
  Weniger}]{Bartels:2015aea}
Bartels, R., Krishnamurthy, S., \& Weniger, C. 2016, Phys. Rev. Lett., 116,
  051102, \dodoi{10.1103/PhysRevLett.116.051102}

\bibitem[{{Bartels} {et~al.}(2018){Bartels}, {Storm}, {Weniger}, \&
  {Calore}}]{2018NatAs...2..819B}
{Bartels}, R., {Storm}, E., {Weniger}, C., \& {Calore}, F. 2018, Nature
  Astronomy, 2, 819, \dodoi{10.1038/s41550-018-0531-z}

\bibitem[{{Bhatt} {et~al.}(2020){Bhatt}, {Sushch}, {Pohl}, {Fedynitch}, {Das},
  {Brose}, {Plotko}, \& {Meyer}}]{2020APh...12302490B}
{Bhatt}, M., {Sushch}, I., {Pohl}, M., {et~al.} 2020, Astroparticle Physics,
  123, 102490, \dodoi{10.1016/j.astropartphys.2020.102490}

\bibitem[{{Bissantz} {et~al.}(2003){Bissantz}, {Englmaier}, \&
  {Gerhard}}]{2003MNRAS.340..949B}
{Bissantz}, N., {Englmaier}, P., \& {Gerhard}, O. 2003, \mnras, 340, 949,
  \dodoi{10.1046/j.1365-8711.2003.06358.x}

\bibitem[{{Bland-Hawthorn} \& {Gerhard}(2016)}]{BlandHawthornandGerhard}
{Bland-Hawthorn}, J., \& {Gerhard}, O. 2016, \araa, 54, 529,
  \dodoi{10.1146/annurev-astro-081915-023441}

\bibitem[{{Blumenthal} \& {Gould}(1970)}]{1970RvMP...42..237B}
{Blumenthal}, G.~R., \& {Gould}, R.~J. 1970, Reviews of Modern Physics, 42,
  237, \dodoi{10.1103/RevModPhys.42.237}

\bibitem[{{Buschmann} {et~al.}(2020){Buschmann}, {Rodd}, {Safdi}, {Chang},
  {Mishra-Sharma}, {Lisanti}, \& {Macias}}]{Buschmann:2020adf}
{Buschmann}, M., {Rodd}, N.~L., {Safdi}, B.~R., {et~al.} 2020, \prd, 102,
  023023, \dodoi{10.1103/PhysRevD.102.023023}

\bibitem[{{Calabretta} {et~al.}(2014){Calabretta}, {Staveley-Smith}, \&
  {Barnes}}]{2014PASA...31....7C}
{Calabretta}, M.~R., {Staveley-Smith}, L., \& {Barnes}, D.~G. 2014, \pasa, 31,
  e007, \dodoi{10.1017/pasa.2013.36}

\bibitem[{Calore {et~al.}(2021)Calore, Donato, \& Manconi}]{Calore:2021jvg}
Calore, F., Donato, F., \& Manconi, S. 2021, Phys. Rev. Lett., 127, 161102,
  \dodoi{10.1103/PhysRevLett.127.161102}

\bibitem[{{Chang} {et~al.}(2020){Chang}, {Mishra-Sharma}, {Lisanti},
  {Buschmann}, {Rodd}, \& {Safdi}}]{Chang2020}
{Chang}, L.~J., {Mishra-Sharma}, S., {Lisanti}, M., {et~al.} 2020, Phys. Rev.
  D, 101, 023014, \dodoi{10.1103/PhysRevD.101.023014}

\bibitem[{{Chrob{\'a}kov{\'a}} {et~al.}(2020){Chrob{\'a}kov{\'a}},
  {L{\'o}pez-Corredoira}, {Sylos Labini}, {Wang}, \&
  {Nagy}}]{2020A&A...642A..95C}
{Chrob{\'a}kov{\'a}}, {\v{Z}}., {L{\'o}pez-Corredoira}, M., {Sylos Labini}, F.,
  {Wang}, H.~F., \& {Nagy}, R. 2020, \aap, 642, A95,
  \dodoi{10.1051/0004-6361/202038736}

\bibitem[{Coleman {et~al.}(2020)Coleman, Paterson, Gordon, Macias, \&
  Ploeg}]{Coleman:2019kax}
Coleman, B., Paterson, D., Gordon, C., Macias, O., \& Ploeg, H. 2020, Mon. Not.
  Roy. Astron. Soc., 495, 3350, \dodoi{10.1093/mnras/staa1281}

\bibitem[{Daylan {et~al.}(2016)Daylan, Finkbeiner, Hooper, Linden, Portillo,
  Rodd, \& Slatyer}]{Daylan:2014rsa}
Daylan, T., Finkbeiner, D.~P., Hooper, D., {et~al.} 2016, Phys. Dark Univ., 12,
  1, \dodoi{10.1016/j.dark.2015.12.005}

\bibitem[{Di~Mauro(2021)}]{DiMauro:2021raz}
Di~Mauro, M. 2021, Phys. Rev. D, 103, 063029,
  \dodoi{10.1103/PhysRevD.103.063029}

\bibitem[{{Dickey} \& {Lockman}(1990)}]{1990ARA&A..28..215D}
{Dickey}, J.~M., \& {Lockman}, F.~J. 1990, \araa, 28, 215,
  \dodoi{10.1146/annurev.aa.28.090190.001243}

\bibitem[{{Draine}(2011)}]{Draine2011}
{Draine}, B.~T. 2011, {Physics of the Interstellar and Intergalactic Medium}

\bibitem[{{Federici} {et~al.}(2015){Federici}, {Pohl}, {Telezhinsky},
  {Wilhelm}, \& {Dwarkadas}}]{2015A&A...577A..12F}
{Federici}, S., {Pohl}, M., {Telezhinsky}, I., {Wilhelm}, A., \& {Dwarkadas},
  V.~V. 2015, \aap, 577, A12, \dodoi{10.1051/0004-6361/201424947}

\bibitem[{{Gibson} {et~al.}(2005{\natexlab{a}}){Gibson}, {Taylor}, {Higgs},
  {Brunt}, \& {Dewdney}}]{2005ApJ...626..214G}
{Gibson}, S.~J., {Taylor}, A.~R., {Higgs}, L.~A., {Brunt}, C.~M., \& {Dewdney},
  P.~E. 2005{\natexlab{a}}, \apj, 626, 214, \dodoi{10.1086/429871}

\bibitem[{{Gibson} {et~al.}(2005{\natexlab{b}}){Gibson}, {Taylor}, {Higgs},
  {Brunt}, \& {Dewdney}}]{2005ApJ...626..195G}
---. 2005{\natexlab{b}}, \apj, 626, 195, \dodoi{10.1086/429870}

\bibitem[{{Goodenough} \& {Hooper}(2009)}]{2009arXiv0910.2998G}
{Goodenough}, L., \& {Hooper}, D. 2009, arXiv e-prints, arXiv:0910.2998.
\newblock \doarXiv{0910.2998}

\bibitem[{{Grenier} {et~al.}(2005){Grenier}, {Casandjian}, \&
  {Terrier}}]{Grenier2005}
{Grenier}, I.~A., {Casandjian}, J.-M., \& {Terrier}, R. 2005, Science, 307,
  1292, \dodoi{10.1126/science.1106924}

\bibitem[{{HI4PI Collaboration} {et~al.}(2016){HI4PI Collaboration}, {Ben
  Bekhti}, {Fl{\"o}er}, {Keller}, {Kerp}, {Lenz}, {Winkel}, {Bailin},
  {Calabretta}, {Dedes}, {Ford}, {Gibson}, {Haud}, {Janowiecki}, {Kalberla},
  {Lockman}, {McClure-Griffiths}, {Murphy}, {Nakanishi}, {Pisano}, \&
  {Staveley-Smith}}]{2016A&A...594A.116H}
{HI4PI Collaboration}, {Ben Bekhti}, N., {Fl{\"o}er}, L., {et~al.} 2016, \aap,
  594, A116, \dodoi{10.1051/0004-6361/201629178}

\bibitem[{{Hooper} \& {Goodenough}(2011)}]{2011PhLB..697..412H}
{Hooper}, D., \& {Goodenough}, L. 2011, Physics Letters B, 697, 412,
  \dodoi{10.1016/j.physletb.2011.02.029}

\bibitem[{{J{\'o}hannesson} {et~al.}(2018{\natexlab{a}}){J{\'o}hannesson},
  {Porter}, \& {Moskalenko}}]{2018ApJ...856...45J}
{J{\'o}hannesson}, G., {Porter}, T.~A., \& {Moskalenko}, I.~V.
  2018{\natexlab{a}}, \apj, 856, 45, \dodoi{10.3847/1538-4357/aab26e}

\bibitem[{{J{\'o}hannesson} {et~al.}(2018{\natexlab{b}}){J{\'o}hannesson},
  {Porter}, \& {Moskalenko}}]{Johannesson:2018bit}
---. 2018{\natexlab{b}}, \apj, 856, 45, \dodoi{10.3847/1538-4357/aab26e}

\bibitem[{{Kalberla} \& {Haud}(2015)}]{2015A&A...578A..78K}
{Kalberla}, P.~M.~W., \& {Haud}, U. 2015, \aap, 578, A78,
  \dodoi{10.1051/0004-6361/201525859}

\bibitem[{{Kalberla} {et~al.}(1980){Kalberla}, {Mebold}, \&
  {Reich}}]{1980A&A....82..275K}
{Kalberla}, P.~M.~W., {Mebold}, U., \& {Reich}, W. 1980, \aap, 82, 275

\bibitem[{{Kerp} {et~al.}(2011){Kerp}, {Winkel}, {Ben Bekhti}, {Fl{\"o}er}, \&
  {Kalberla}}]{2011AN....332..637K}
{Kerp}, J., {Winkel}, B., {Ben Bekhti}, N., {Fl{\"o}er}, L., \& {Kalberla},
  P.~M.~W. 2011, Astronomische Nachrichten, 332, 637,
  \dodoi{10.1002/asna.201011548}

\bibitem[{{Leane} \& {Slatyer}(2019)}]{LeaneSlatyer2019}
{Leane}, R.~K., \& {Slatyer}, T.~R. 2019, arXiv e-prints, arXiv:1904.08430.
\newblock \doarXiv{1904.08430}

\bibitem[{{Leane} \& {Slatyer}(2020{\natexlab{a}})}]{LeaneSlatyer2020}
---. 2020{\natexlab{a}}, Phys. Rev. D, 102, 063019,
  \dodoi{10.1103/PhysRevD.102.063019}

\bibitem[{{Leane} \& {Slatyer}(2020{\natexlab{b}})}]{Leane2020}
---. 2020{\natexlab{b}}, \prl, 125, 121105,
  \dodoi{10.1103/PhysRevLett.125.121105}

\bibitem[{Lee {et~al.}(2016)Lee, Lisanti, Safdi, Slatyer, \& Xue}]{Lee:2015fea}
Lee, S.~K., Lisanti, M., Safdi, B.~R., Slatyer, T.~R., \& Xue, W. 2016, Phys.
  Rev. Lett., 116, 051103, \dodoi{10.1103/PhysRevLett.116.051103}

\bibitem[{{List} {et~al.}(2021){List}, {Rodd}, \& {Lewis}}]{List:2021aer}
{List}, F., {Rodd}, N.~L., \& {Lewis}, G.~F. 2021, arXiv e-prints,
  arXiv:2107.09070.
\newblock \doarXiv{2107.09070}

\bibitem[{{Macias} {et~al.}(2018){Macias}, {Gordon}, {Crocker}, {Coleman},
  {Paterson}, {Horiuchi}, \& {Pohl}}]{2018NatAs...2..387M}
{Macias}, O., {Gordon}, C., {Crocker}, R.~M., {et~al.} 2018, Nature Astronomy,
  2, 387, \dodoi{10.1038/s41550-018-0414-3}

\bibitem[{{Macias} {et~al.}(2019){Macias}, {Horiuchi}, {Kaplinghat}, {Gordon},
  {Crocker}, \& {Nataf}}]{Macias:2019omb}
{Macias}, O., {Horiuchi}, S., {Kaplinghat}, M., {et~al.} 2019, \jcap, 2019,
  042, \dodoi{10.1088/1475-7516/2019/09/042}

\bibitem[{{Macias} {et~al.}(2021){Macias}, {van Leijen}, {Song}, {Ando},
  {Horiuchi}, \& {Crocker}}]{Macias:2021boz}
{Macias}, O., {van Leijen}, H., {Song}, D., {et~al.} 2021, \mnras, 506, 1741,
  \dodoi{10.1093/mnras/stab1450}

\bibitem[{{McClure-Griffiths} {et~al.}(2009){McClure-Griffiths}, {Pisano},
  {Calabretta}, {Ford}, {Lockman}, {Staveley-Smith}, {Kalberla}, {Bailin},
  {Dedes}, {Janowiecki}, {Gibson}, {Murphy}, {Nakanishi}, \&
  {Newton-McGee}}]{2009ApJS..181..398M}
{McClure-Griffiths}, N.~M., {Pisano}, D.~J., {Calabretta}, M.~R., {et~al.}
  2009, \apjs, 181, 398, \dodoi{10.1088/0067-0049/181/2/398}

\bibitem[{Mertsch \& Phan(2022)}]{mertsch2022bayesian}
Mertsch, P., \& Phan, V. H.~M. 2022, Bayesian inference of three-dimensional
  gas maps: II. Galactic HI.
\newblock \doarXiv{2202.02341}

\bibitem[{{Mertsch} \& {Vittino}(2020)}]{2020arXiv201215770M}
{Mertsch}, P., \& {Vittino}, A. 2020, arXiv e-prints, arXiv:2012.15770.
\newblock \doarXiv{2012.15770}

\bibitem[{{Mishra-Sharma} \& {Cranmer}(2021)}]{Mishra-Sharma:2021oxe}
{Mishra-Sharma}, S., \& {Cranmer}, K. 2021, arXiv e-prints, arXiv:2110.06931.
\newblock \doarXiv{2110.06931}

\bibitem[{{Nakanishi} \& {Sofue}(2006)}]{2006PASJ...58..847N}
{Nakanishi}, H., \& {Sofue}, Y. 2006, \pasj, 58, 847,
  \dodoi{10.1093/pasj/58.5.847}

\bibitem[{Nishiyama {et~al.}(2013)Nishiyama, Yasui, Nagata, Yoshikawa,
  Uchiyama, Schdel, Hatano, Sato, Sugitani, Suenaga, Kwon, \&
  Tamura}]{Nishiyama2015}
Nishiyama, S., Yasui, K., Nagata, T., {et~al.} 2013, ApJ. Lett., 769, L28.
\newblock \url{http://stacks.iop.org/2041-8205/769/i=2/a=L28}

\bibitem[{{Pettitt} {et~al.}(2014){Pettitt}, {Dobbs}, {Acreman}, \&
  {Price}}]{2014MNRAS.444..919P}
{Pettitt}, A.~R., {Dobbs}, C.~L., {Acreman}, D.~M., \& {Price}, D.~J. 2014,
  \mnras, 444, 919, \dodoi{10.1093/mnras/stu1075}

\bibitem[{{Planck Collaboration} {et~al.}(2016){Planck Collaboration},
  {Aghanim}, {Ashdown}, {Aumont}, {Baccigalupi}, {Ballardini}, {Banday},
  {Barreiro}, {Bartolo}, {Basak}, {Benabed}, {Bernard}, {Bersanelli},
  {Bielewicz}, {Bonavera}, {Bond}, {Borrill}, {Bouchet}, {Boulanger},
  {Burigana}, {Calabrese}, {Cardoso}, {Carron}, {Chiang}, {Colombo}, {Comis},
  {Couchot}, {Coulais}, {Crill}, {Curto}, {Cuttaia}, {de Bernardis}, {de
  Zotti}, {Delabrouille}, {Di Valentino}, {Dickinson}, {Diego}, {Dor{\'e}},
  {Douspis}, {Ducout}, {Dupac}, {Dusini}, {Elsner}, {En{\ss}lin}, {Eriksen},
  {Falgarone}, {Fantaye}, {Finelli}, {Forastieri}, {Frailis}, {Fraisse},
  {Franceschi}, {Frolov}, {Galeotta}, {Galli}, {Ganga}, {G{\'e}nova-Santos},
  {Gerbino}, {Ghosh}, {Giraud-H{\'e}raud}, {Gonz{\'a}lez-Nuevo}, {G{\'o}rski},
  {Gruppuso}, {Gudmundsson}, {Hansen}, {Helou}, {Henrot-Versill{\'e}},
  {Herranz}, {Hivon}, {Huang}, {Jaffe}, {Jones}, {Keih{\"a}nen}, {Keskitalo},
  {Kiiveri}, {Kisner}, {Krachmalnicoff}, {Kunz}, {Kurki-Suonio}, {Lamarre},
  {Langer}, {Lasenby}, {Lattanzi}, {Lawrence}, {Le Jeune}, {Levrier}, {Lilje},
  {Lilley}, {Lindholm}, {L{\'o}pez-Caniego}, {Ma}, {Mac{\'\i}as-P{\'e}rez},
  {Maggio}, {Maino}, {Mandolesi}, {Mangilli}, {Maris}, {Martin},
  {Mart{\'\i}nez-Gonz{\'a}lez}, {Matarrese}, {Mauri}, {McEwen}, {Melchiorri},
  {Mennella}, {Migliaccio}, {Miville-Desch{\^e}nes}, {Molinari}, {Moneti},
  {Montier}, {Morgante}, {Moss}, {Natoli}, {Oxborrow}, {Pagano}, {Paoletti},
  {Patanchon}, {Perdereau}, {Perotto}, {Pettorino}, {Piacentini},
  {Plaszczynski}, {Polastri}, {Polenta}, {Puget}, {Rachen}, {Racine},
  {Reinecke}, {Remazeilles}, {Renzi}, {Rocha}, {Rosset}, {Rossetti}, {Roudier},
  {Rubi{\~n}o-Mart{\'\i}n}, {Ruiz-Granados}, {Salvati}, {Sandri}, {Savelainen},
  {Scott}, {Sirignano}, {Sirri}, {Soler}, {Spencer}, {Suur-Uski}, {Tauber},
  {Tavagnacco}, {Tenti}, {Toffolatti}, {Tomasi}, {Tristram}, {Trombetti},
  {Valiviita}, {Van Tent}, {Vielva}, {Villa}, {Vittorio}, {Wandelt}, {Wehus},
  {Zacchei}, \& {Zonca}}]{Planck_dust}
{Planck Collaboration}, {Aghanim}, N., {Ashdown}, M., {et~al.} 2016, \aap, 596,
  A109, \dodoi{10.1051/0004-6361/201629022}

\bibitem[{{Pohl} {et~al.}(2008){Pohl}, {Englmaier}, \&
  {Bissantz}}]{2008ApJ...677..283P}
{Pohl}, M., {Englmaier}, P., \& {Bissantz}, N. 2008, \apj, 677, 283,
  \dodoi{10.1086/529004}

\bibitem[{{Porter} {et~al.}(2017){Porter}, {J{\'o}hannesson}, \&
  {Moskalenko}}]{Porter:2017vaa}
{Porter}, T.~A., {J{\'o}hannesson}, G., \& {Moskalenko}, I.~V. 2017, \apj, 846,
  67, \dodoi{10.3847/1538-4357/aa844d}

\bibitem[{Read {et~al.}(2016)Read, Agertz, \& Collins}]{Read:2015sta}
Read, J.~I., Agertz, O., \& Collins, M. L.~M. 2016, Mon. Not. Roy. Astron.
  Soc., 459, 2573, \dodoi{10.1093/mnras/stw713}

\bibitem[{{Reich} {et~al.}(2001){Reich}, {Testori}, \&
  {Reich}}]{2001A&A...376..861R}
{Reich}, P., {Testori}, J.~C., \& {Reich}, W. 2001, \aap, 376, 861,
  \dodoi{10.1051/0004-6361:20011000}

\bibitem[{{Selig} {et~al.}(2015){Selig}, {Vacca}, {Oppermann}, \&
  {En{\ss}lin}}]{2015A&A...581A.126S}
{Selig}, M., {Vacca}, V., {Oppermann}, N., \& {En{\ss}lin}, T.~A. 2015, \aap,
  581, A126, \dodoi{10.1051/0004-6361/201425172}

\bibitem[{Slatyer(2021)}]{Slatyer:2021qgc}
Slatyer, T.~R. 2021, in {Les Houches summer school on Dark Matter}.
\newblock \doarXiv{2109.02696}

\bibitem[{{Sofue}(2015)}]{2015PASJ...67...75S}
{Sofue}, Y. 2015, \pasj, 67, 75, \dodoi{10.1093/pasj/psv042}

\bibitem[{{Song} {et~al.}(2019){Song}, {Macias}, \& {Horiuchi}}]{Song:2019nrx}
{Song}, D., {Macias}, O., \& {Horiuchi}, S. 2019, \prd, 99, 123020,
  \dodoi{10.1103/PhysRevD.99.123020}

\bibitem[{{Su} {et~al.}(2010){Su}, {Slatyer}, \&
  {Finkbeiner}}]{2010ApJ...724.1044S}
{Su}, M., {Slatyer}, T.~R., \& {Finkbeiner}, D.~P. 2010, \apj, 724, 1044,
  \dodoi{10.1088/0004-637X/724/2/1044}

\bibitem[{{Testori} {et~al.}(2001){Testori}, {Reich}, {Bava}, {Colomb},
  {Hurrel}, {Larrarte}, {Reich}, \& {Sanz}}]{2001A&A...368.1123T}
{Testori}, J.~C., {Reich}, P., {Bava}, J.~A., {et~al.} 2001, \aap, 368, 1123,
  \dodoi{10.1051/0004-6361:20010088}

\bibitem[{{Winkel} {et~al.}(2016){Winkel}, {Kerp}, {Fl{\"o}er}, {Kalberla},
  {Ben Bekhti}, {Keller}, \& {Lenz}}]{2016A&A...585A..41W}
{Winkel}, B., {Kerp}, J., {Fl{\"o}er}, L., {et~al.} 2016, \aap, 585, A41,
  \dodoi{10.1051/0004-6361/201527007}

\bibitem[{{Wolleben}(2007)}]{Wolleben:2007}
{Wolleben}, M. 2007, Astrophys.~J., 664, 349, \dodoi{10.1086/518711}

\end{thebibliography}
\bibliographystyle{aasjournal}

\appendix

\section{Gamma-ray Analysis Procedure}
\label{sec:fermianalysis}

We used a similar pipeline to that introduced in \citet[e.g.,][]{Macias:2019omb,Abazajian:2020tww}.
In particular, we fitted our ROI model (see Appendix \ref{sec:astrotemplates}) to the data using a bin-by-bin analysis procedure in which we maximized the likelihood function separately at each individual energy bin. We obtained the bin fluxes for each template by assuming a simple power-law, $dN/dE=N_0 E^{-2}$, {within each bin, but not bin-to-bin. Instead, we freely vary} the bin-wise normalisation $N_0$ of all the templates in the fits. An advantage of using a bin-by-bin procedure over a broad-band analysis is that, with the former, 
there is little or no need to make assumptions about the spectral shape of a new template, rather 
the template spectrum is obtained in a completely data-driven way. We note in passing that the fitting was done with the \textsc{pylikelihood} routine, the standard maximum-likelihood method in \textsc{Fermitools}.
In our analysis, we used the bin-by-bin method to evaluate the best-fit spectral values, and the statistical significance for each new source added in our ROI model. An advantage of using \texttt{Fermitools} for our fits, is that it rigorously accounts for the point spread function of the LAT. 
In order to evaluate the statistical significance of a new template we compute the $\Delta \mathrm{TS}$ for the full energy range as shown in Eq.~2.5 of \cite{Macias:2019omb}. Note that since the normalization of the sources are varied independently at each energy bin, we need to use the mixture distribution, explained in \cite{Macias:2019omb}, to correctly compute the statistical significance of a new source. In doing so, we account for the number of degrees of freedom for a new extended source, which is the same as the number of energy bins.

\section{Astrophysical Templates}
\label{sec:astrotemplates}

\subsection*{Hadronic and Bremsstrahlung gamma rays}
The dominant contributions to the gamma-ray emission within our ROI are hadronic and bremsstrahlung emission resulting from the interaction of Galactic cosmic-ray protons and electrons with interstellar gas.
Since both of these components are spatially correlated with the distribution of interstellar gas, we model them in a data-driven way. Namely, we include templates of H$I$, H$_2$, and dust correction maps in our ROI model and then reconstruct their spectra using the bin-by-bin fitting procedure explained in Appendix~\ref{sec:fermianalysis}. The H$_2$ maps are the same as those in~\citet{2018NatAs...2..387M}, whereas the hydrodynamic H$I$ and dust correction templates are updated. The ring subdivision of the gas maps allows to account for the radial evolution of cosmic ray density, and the small width of the energy bins permits to recover their respective gamma-ray spectra.  

For our Base model, we selected the hydrodynamic H$I$ and dust correction maps denoted as ``best $T_{\rm exc}$'', as these provide the statistically most favoured model of the line spectra of atomic hydrogen.

\subsection*{Dust correction templates}
Molecular hydrogen that is not well mixed with carbon monoxide will not be traced by the CO 2.6 mm emission. Furthermore, 
assuming a constant atomic hydrogen spin temperature
along a particular line of sight can give an incorrect estimate of column density. 
To correct for these deficiencies we included 
dust correction templates based on the methods used by \cite{Fermi-LAT:4FGL}\footnote{See also \url{https://fermi.gsfc.nasa.gov/ssc/data/analysis/software/aux/4fgl/Galactic_Diffuse_Emission_Model_for_the_4FGL_Catalog_Analysis.pdf}.}.
  Infrared thermal emission from dust provides an alternative method of tracing hydrogen gas in the Milky Way \citep{Grenier2005}. The correction templates are obtained by subtracting the components of the dust emission that are correlated with the gas already traced by 21 cm and 2.6 mm emission. 

\begin{figure*}
\includegraphics[width=0.49\linewidth]{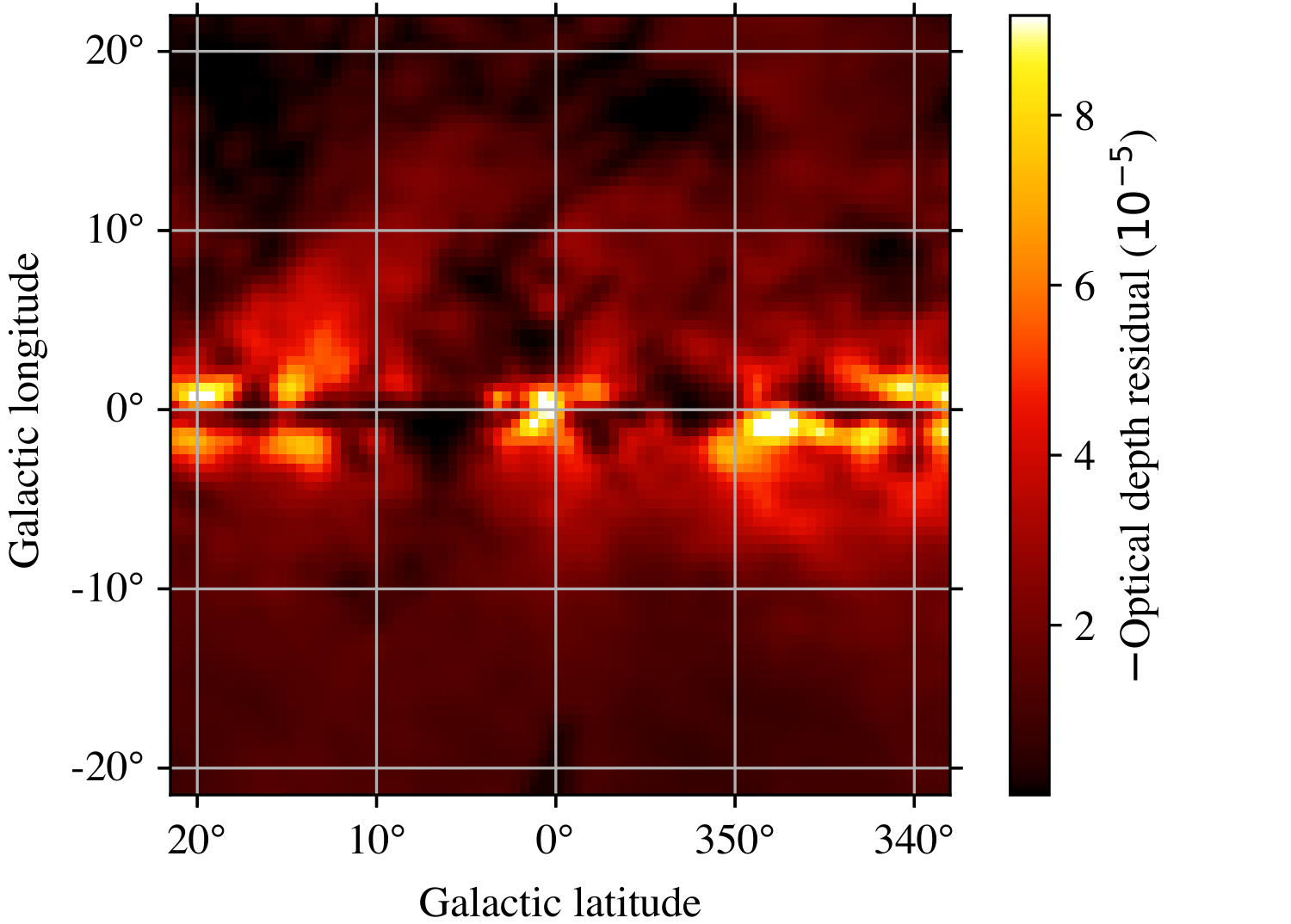} 
 \includegraphics[width=0.49\linewidth]{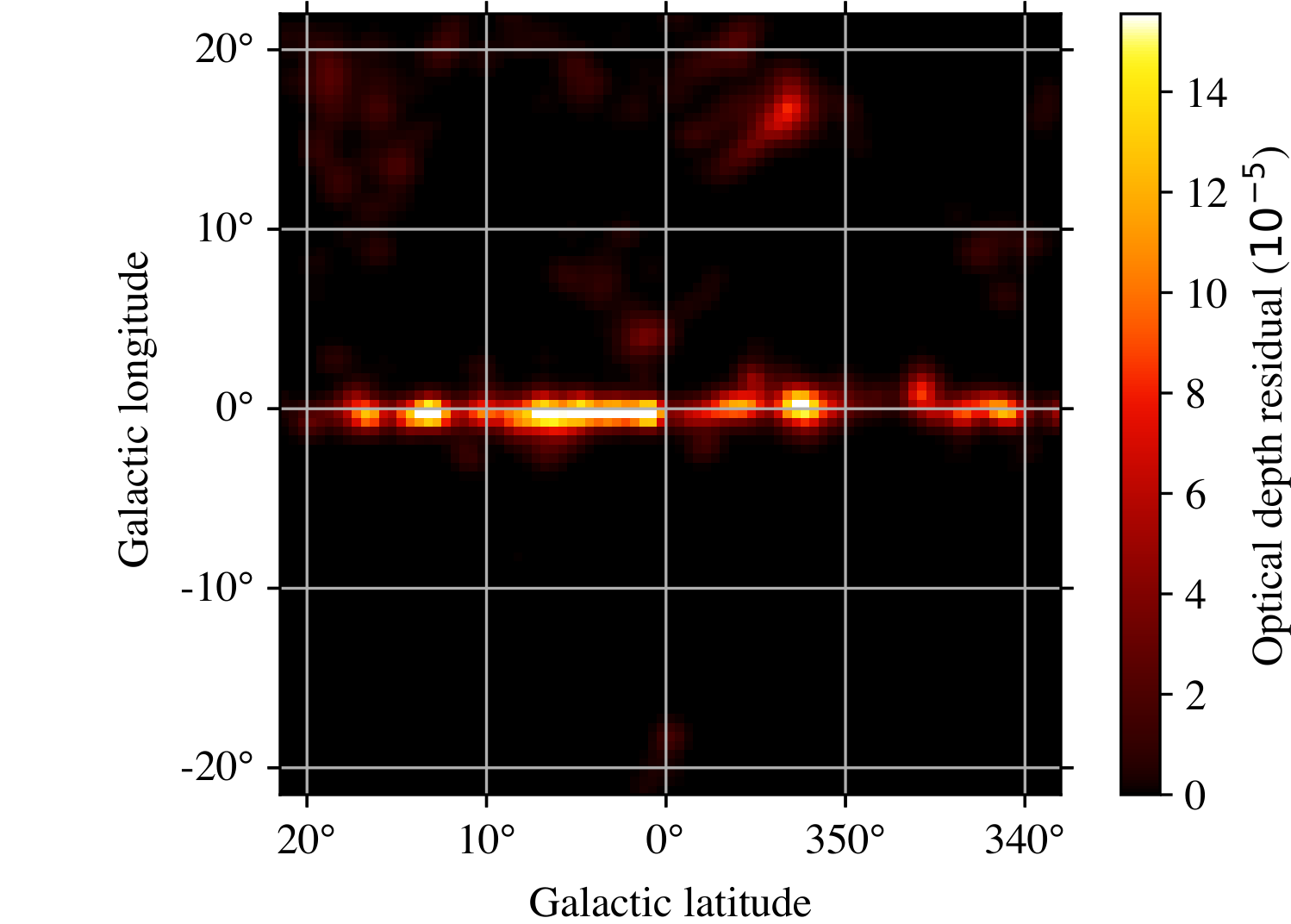} 
 \caption{Dust residual maps. On the left we have plotted the negative residuals (multiplied by $-1$) and on the right the positive residuals. For display purposes the images have been smoothed with a Gaussian filter of radius $0.5^\circ$, and the colorbar has been chosen to encompass 99.5\% of intensity values.}
 \label{fig:dust_residuals}
 \end{figure*}

We applied this method to the Planck dust optical depth map\footnote{COM\_CompMap\_Dust-GNILC-Model-Opacity\_2048\_R2.01.fits, \cite{Planck_dust}}. After subtracting the components of the Planck dust optical depth map that were linearly correlated with our estimated atomic and molecular hydrogen gas maps, the residuals were separated into positive and negative components. The positive residuals physically represent hydrogen that is not traced by the relevant emission, known as the dark neutral medium, or an over estimation of the atomic hydrogen spin temperature. Negative residuals represent an underestimation of the spin temperature. The results are displayed in Fig.~\ref{fig:dust_residuals}.

\subsection*{Inverse Compton emission}
We similarly need a model for diffuse, inverse Compton (IC) emission, the second largest source of background, and here we choose the six-ring IC model introduced in~\cite{Abazajian:2020tww},  because it allows to account for modelling uncertainties such as a potential new central source of electron or possible bias introduced by assumptions on the normalization and shape of the interstellar radiation field (ISRF) and the electron injection spectra.

We constructed the IC maps using \texttt{GALPROP~V56} and the propagation parameter setup  SA50,  shown in Table 5 of \citet{Johannesson:2018bit}. It should be noted that these IC maps are based on 3D models for the ISRF~\citep{Porter:2017vaa}, and hence they do not need the simplifying assumptions that were required in previous versions of the \texttt{GALPROP} code.
\subsection*{Low-latitude Fermi bubbles}
The Fermi bubbles (FB) are one of the strongest sources of fore-/background emission in our sky region.
The FB are themselves defined as highly statistically significant and spatially coherent gamma-ray residuals, whose spectra are well described by a simple power-law with a relatively hard slope, $s\simeq -1.9$, in the energy range of our analysis~\citep{2010ApJ...724.1044S, Fermi-LAT:2014sfa}.
Here, we use an improved version~\citep{Macias:2019omb} of the FB  template constructed using a spectral component analysis by the {\it Fermi} Collaboration~\citep{Fermi-LAT:2014sfa}.

\subsection*{Point Sources} 

We modelled gamma-ray point sources using the 4FGL \citep{Fermi-LAT:4FGL}. 
Specifically, we assumed the release \texttt{gll$\_$psc$\_$v20.fit}, which contains 487 gamma-ray point sources in our ROI.
Unfortunately, varying the normalization of all these point sources at once in a maximum-likelihood run is very challenging, and so we opted for following the hybrid-modelling procedure implemented in~\cite{Macias:2019omb}. In particular, we floated the normalization of each of the 120 brightest point sources in our RoI, and for the remaining 367 sources we constructed a point source population template assuming the flux ratios reported in the 4FGL. The population template was included in the fits with its overall normalization free to vary at each energy bin.
This is a reasonable simplification given that our data selection cuts are the same as those in the 4FGL.

\subsection*{GCE Templates}

Detailed descriptions of the templates used to model the GCE are given in Appendix B of~\cite{Abazajian:2020tww} (and references therein). In summary, we model the GCE signal with maps tracing the distribution of stellar mass in the Galactic bulge, or with maps describing the distribution of Galactic DM. For the bulge stars, we included two independent templates: the ``Boxy Bulge'' (BB) model proposed in~\citet{Coleman:2019kax}, and the observational ``Nuclear Bulge'' (NB) map presented in~\cite{Nishiyama2015}. For the DM distribution, we used a cuspy profile, given by a Navarro-Frenk-White (NFW) model with slope $\gamma=1.2$, and a cored profile, given by a Read function with $\gamma=1.0$ and core radius 1 kpc~\citep{Read:2015sta}. Furthermore, we considered ellipsoidal versions of these. For full details of our profile choices see Fig.~3 of \citet{Abazajian:2020tww} and text therein.

\subsection*{Other Standard Templates}

Additional extended sources considered in our analysis include Loop I~\citep{Wolleben:2007}, maps for the \textit{Sun} and the \textit{Moon} tailor-made for our data selection cuts, extracted from the 4FGL, and an isotropic gamma-ray model (\texttt{iso$_{-}$P8R3$_{-}$ULTRACLEANVETO$_{-}$V2$_{-}$v1.txt}). 

\section{Gamma-ray Spectrum}

\begin{figure}
    \centering
    \includegraphics{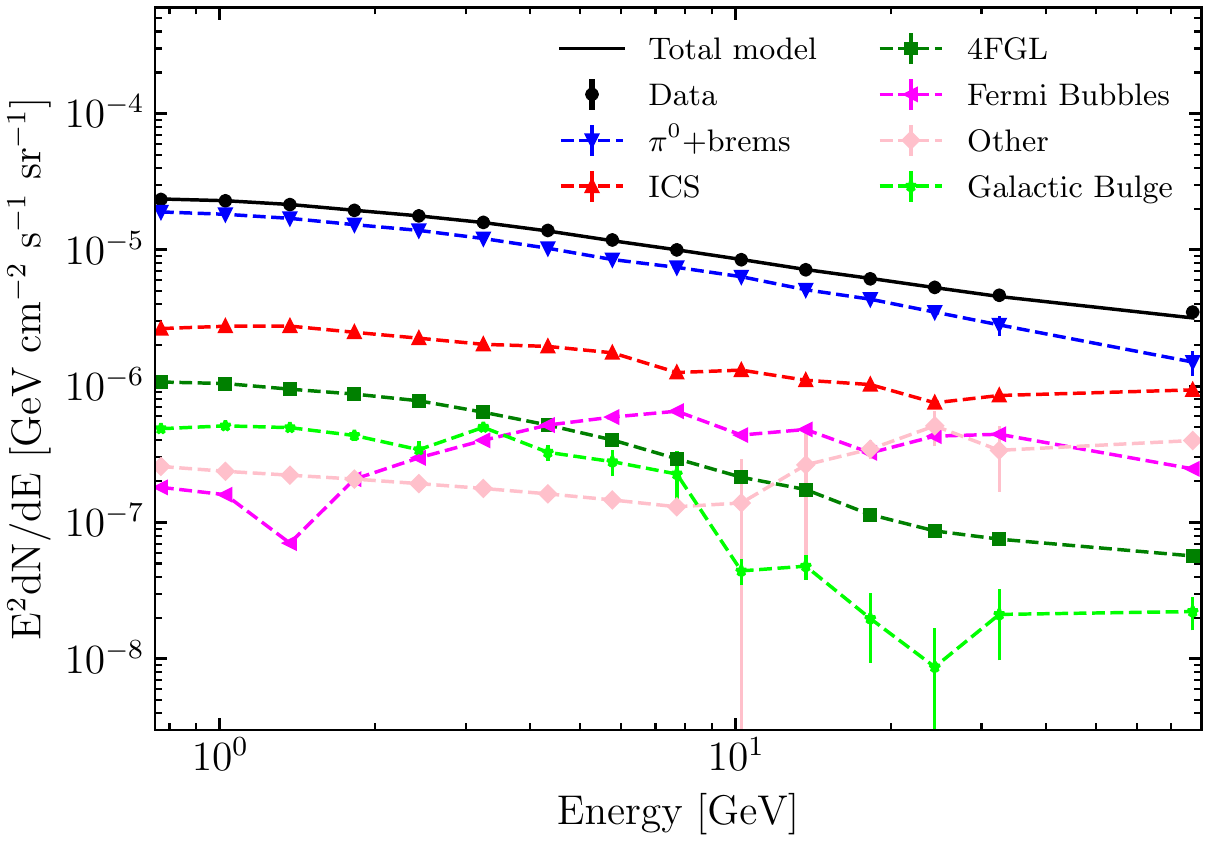}
    \caption{The best-fit spectra for components of the ``Base+NB+BB'' model, cf. Table~\ref{tab:loglike-values}. The Base model includes the H$I$ maps with $T_{\rm exc}$ varying with $l$ and $b$, divided in four concentric rings, 3D inverse Compton (IC) maps divided in six concentric rings, the 4FGL catalog of point sources~\citep{Fermi-LAT:4FGL}, the Fermi Bubbles template, specialized templates for the Sun and Moon, an isotropic emission model, and a geometrical template for Loop I (see Appendix~\ref{sec:astrotemplates}). We have grouped several components together for presentation purposes.  }
    \label{fig:spectra}
 \end{figure}

The spectrum for the best-fitting ``Base+NB+BB'' model (see Table~\ref{tab:loglike-values}) is shown in Fig.\ \ref{fig:spectra}. As in our previous studies~\citep[e.g.,][]{2018NatAs...2..387M,Macias:2019omb,Abazajian:2020tww}, we find that the fitting procedure produces physically sensible spectra. For simplicity, we have thematically grouped the spectra of different templates. For example, the spectra for the H$I$, CO, and dust maps is displayed as ``$\pi^0$+bremss'', the spectra of all the gamma-ray point sources is shown as one single line denoted ``4FGL'', and likewise for the rest of the templates  of our ROI model.

\section{Impact of H\textit{I} systematics on the GCE}
\label{sec:200K}

Previous studies~\citep[e.g.,][]{2018NatAs...2..387M,2018NatAs...2..819B,Macias:2019omb, Abazajian:2020tww} have demonstrated that the GCE signal is better explained by stellar mass templates than DM templates. As stated in the introduction, those studies assumed H$I$ maps with a constant excitation temperature through the Galaxy. In contrast, the fiducial H$I$ maps included in the main pipeline of the present study consider a 
 excitation temperature which varies with longitude and latitude and accounts for the radiation transport in the presence of continuum emission. In this section, we repeat the hierarchical fitting procedure, whose results are summarized in Table~\ref{tab:loglike-values}, except that this time we replace the fiducial H$I$ maps with those for $T_{\rm exc}=200$~K that among all tested constant excitation temperatures provides the best reproduction of the H$I$ line spectra (see Fig.~\ref{fig:noise}).

Table~\ref{tab:SystematicsLogLikes} shows the statistical significance for each of the GCE templates for fixed $T_{\rm exc}=200$~K. To be noted is that they are qualitatively and quantitatively very similar to those obtained with our fiducial ROI model. We again find that the GCE data strongly prefers the stellar-mass templates as a proxy for the GCE morphology. Furthermore, as was found with our Base model and with the ``Alt. Base'', we find that in order to remove any support of the DM hypothesis, it is sufficient to add the NB template to the ``Alt. Base''.  However, as seen in row 12 of Table~\ref{tab:SystematicsLogLikes}, the BB template is still required by the data. We thus include the BB template into the sky model and confirm our negative DM results.

\begin{table}[ht!]
\centering
\begin{tabular}{cccc}
\hline\hline
Baseline model  & Additional source &  $\Delta \mathrm{TS}$  &  Significance \\ 
                &                   &               &   \\\hline
Alt. Base  & Cored ellipsoidal &  1.3         & $0.0\;\sigma$\\
Alt. Base  & Cored             &  2.0         & $0.1\;\sigma$\\
Alt. Base  & BB                &  304.2       & $15.9\;\sigma$\\ 
Alt. Base  & NFW ellipsoidal   &  682.7        & $24.9\;\sigma$\\ 
Alt. Base  & NFW               &  837.9        & $27.8\;\sigma$\\ 
Alt. Base  & NB                & 1753.0       & $41.1\;\sigma$\\ \hline
Alt. Base+NB  & Cored             &   2.2        & $0.1\;\sigma$\\
Alt. Base+NB  & Cored ellipsoidal &   2.4    & $0.1\;\sigma$\\ 
Alt. Base+NB  & NFW ellipsoidal   &  3.5     & $0.2\;\sigma$\\ 
Alt. Base+NB  & NFW               &  5.7     & $0.5\;\sigma$\\ 
Alt. Base+NB  & BB                &  283.2 &$15.3\;\sigma$\\\hline Alt. Base+NB+BB  & Cored ellipsoidal &  0.1    & $0.0\;\sigma$\\ 
Alt. Base+NB+BB  & NFW ellipsoidal   &  0.5   & $0.0\;\sigma$\\ 
Alt. Base+NB+BB  & Cored             &  0.6  & $0.0\;\sigma$\\   
Alt. Base+NB+BB  & NFW               &    2.3    & $0.1\;\sigma$\\ 
\hline\hline
\end{tabular}
\caption{{\bf Statistical significance of the GCE templates assuming H$I$ maps with $T_{\rm exc}=200$ K}. The ``Alternative Base'' model is the same as that shown in Table~\ref{tab:loglike-values}, except that the fiducial H$I$ maps are replaced with those for a constant excitation temperature, $T_{\rm exc}=200$ K.}\label{tab:SystematicsLogLikes}
\end{table}

\end{document}